\documentclass[12pt]{article}
\usepackage{amsmath}
\usepackage{graphicx,psfrag,epsf}
\usepackage{enumerate}
\usepackage{natbib}
\usepackage{url} 
\usepackage{amsthm} 
\usepackage{float} 
\usepackage{import} 
\usepackage{booktabs}
\usepackage[toc, page] {appendix}

\title{Quadratic Funding and Matching Funds Requirements}

\author{ Ricardo A. Pasquini \thanks{Facultad de Ciencias Empresariales, Universidad Austral, Mariano Acosta 1611, B1630FHB Pilar, Buenos Aires. The
views expressed in the paper do not necessarily reflect those of the institutions I am affiliated with. All errors are my own. This research did not receive any specific grant from funding agencies in the public, commercial, or
not-for-profit sectors.  I would like to thank Cecilia Adrogue, Federico De Cristo, Javier García Sanchez, Juan Llach, Maria Jose Murcia, Belen Pagone, Virginia Sarria Allende, Bruno Zapico and two anonymous referees for their comments and suggestions. An earlier version of this paper was circulated under the title "Quadratic Funding under Constrained Matching Funds: Evidence from Gitcoin". \href{mailto:rpasquini@austral.edu.ar}{rpasquini@austral.edu.ar}} }

\newcommand*{\thisdraft}{This version: July 2022} 
\newcommand*{\firstdraft}{First version: September 2020}  

\date{\thisdraft \\ \firstdraft}

\providecommand{\keywords}[1]{\textbf{\textit{Keywords---}} #1}
\providecommand{\jelcodes}[1]{\textbf{\textit{JEL codes---}} #1}

\newtheorem{theorem}{Theorem}[section]

\newtheorem{proposition}[theorem]{Proposition}
\newtheorem{observation}[theorem]{Observation}

\usepackage[breaklinks=true]{hyperref}
\hypersetup{colorlinks=true,%
citecolor=blue,%
filecolor=blue,%
linkcolor=blue,%
urlcolor=blue}
\usepackage{url}

\begin{document}

\def\spacingset#1{\renewcommand{\baselinestretch}%
{#1}\small\normalsize} \spacingset{1}

\maketitle

\begin{abstract}
In this paper we examine the mechanism proposed by Buterin, Hitzig, and Weyl (2019) for public goods financing, particularly regarding its matching funds requirements, related efficiency implications, and incentives for strategic behavior. Then, we use emerging evidence from Gitcoin Grants, to identify stylized facts in contribution giving and test our propositions. Because of its quadratic design, matching funds requirements scale rapidly, particularly by more numerous and equally contributed projects. As a result, matching funds are exhausted early in the funding rounds, and much space remains for social efficiency improvement. Empirically, there is also a tendency by contributors to give small amounts, scattered among multiple projects, which accelerates this process. Among other findings, we also identify a significant amount of reciprocal backing, which could be consistent with the kind of strategic behavior we discuss.
\end{abstract}

    \keywords{quadratic funding, public goods, Gitcoin}

    \jelcodes{D47, D61, D71, H41}

\newpage
\spacingset{1.45} 

\clearpage

\hypertarget{introduction}{%
\section*{Introduction}\label{introduction}}
\addcontentsline{toc}{section}{Introduction}

The efficient provision of public goods is certainly a central issue for
the economy of any community, ecosystem, or society. The economic
literature recognizes that goods such as open-source software,
knowledge, and urban public equipment, are characterized by
non-exclusion and non-rivalry properties. When the allocation of these
goods is decentralized, these characteristics lead to their
under-provision (Samuelson 1954). On the other hand, the centralized
provision of public goods, faces the challenge of identifying the
preferences of the individuals the public good is aimed to benefit
(Clarke 1971; Groves 1973). In addition, in many settings, there is the
additional problem that the planner must choose between a variety of
public goods that could be financed.

Buterin, Hitzig, and Weyl (2019) (BHW) has recently proposed a
decentralized matching funding mechanism for such a setting, that, under
certain conditions, has the property of achieving an ``optimal provision
for an ecosystem of public goods''\footnote{See Buterin, Hitzig, and
  Weyl (2019), pp.~5178.}. The mechanism, known as \emph{quadratic
funding} (QF), involves a sponsor (i.e., donor, or group of sponsors)
matching the contributions of a (decentralized) community of individuals
that support the creation of public good projects. In this mechanism,
total funds to be received by a public good project (i.e., the size of
the public good) results from applying the \emph{quadratic rule} to the
individual contributions (i.e., from taking the sum of the square-roots
of individual contributions and then taking the square of this value)
and using the sponsor funds to reach the resulting levels.\footnote{See
  Section \ref{QFrule} of this paper.} As the mechanism pays projects
additional funding in proportion to funds committed by other sources, it
also resembles other \emph{matching funds} mechanisms observed, for
example in philanthropic giving, public infrastructure funding, and
public-private startup funding.\footnote{See, for instance, Andreoni
  (2006), Huck and Rasul (2011) and Baker, Payne, and Smart (1999).}

The ability of this innovative mechanism to achieve social optimality
through a decentralized arrangement immediately attracted the attention
of both academics and practitioners, which have put the mechanism
into real applications. Probably the most notable of these is
\emph{Gitcoin Grants} (\url{https://gitcoin.co/grants/}), a financing
platform for open-source software projects related to the Ethereum
blockchain ecosystem. Gitcoin regularly uses the QF mechanism to
allocate grants, and has been a center for discussion and dissemination
of the mechanism.

One of the characteristics of the mechanism, as we discuss in this
paper, is related to the fact that pools of funds provided by donors
will be typically limited in relation to those needed to reach socially
optimal allocations. In practice, the total funding that a project
\emph{should} receive according to the QF rule will be greater than the
matching funds available in the donor pool. In other words, the
mechanism will be typically subject to a limited pool of matching funds
constraint. It therefore seems important to examine the properties of
this mechanism in such a scenario. How much is efficiency compromised by
the donors' funding constraint? What would be an optimal allocation rule
under limited funds and to what extent is it being met by the mechanism?

A second dimension of interest is related to possible strategic behavior
from contributors, that might deteriorate the outcomes of the mechanism.
BHW have pointed out forms of collusion and fraud as potential
vulnerabilities and put forward ideas on their scope. Under what
conditions does the mechanism incentivize strategic behavior by
contributors? Into what degree behavior consistent with such incentives
is observed?

Our aim is to further explore these questions at both the theoretical
and empirical levels. Using emerging evidence from Gitcoin Grants, we
will explore stylized facts related to contributors' behavior and to the
outcomes of funding rounds.

We start, in Section \ref{QFrule}, by examining the question of what
determines the size of the pool of matching funds to achieve optimality
in the BHW sense.\footnote{See Buterin, Hitzig, and Weyl (2019),
  Proposition 3 ``Optimality of Quadratic Finance'', pp.~5175.} We note
that required funds increase non-linearly with the number of
contributors, and as result, any given pool of matching funds will be
rapidly exhausted in most real applications. In addition, we will also
note that required funds increase as contributors' correlate their
investment allocations across projects, an observation that could be
significant in terms of platform design (e.g., designs that might induce
correlations via behavioral effects).

This leads to the question about the efficiency of the mechanism under
limited matching funds. In their paper, BHW recognize that ``even the
wealthiest philanthropists do not have infinite funds and, thus, cannot
simply agree to finance arbitrarily large deficits''.\footnote{See
  Buterin, Hitzig, and Weyl (2019), pp.~5178. BHW also notice that
  requiring contributors to finance any ``\emph{deficits''} (i.e.,
  matching funds required by the QF rule in excess of actual
  contributions) the ``QF mechanism does not yield efficiency''.} Then,
as an alternative, they discuss a variant on the QF mechanism (i.e., the
``Capital Constrained QF Mechanism''), which limits the funding promised
by the mechanism.\footnote{BHW call this variant the ``Capital
  Constrained Quadratic Finance Mechanism''. See BHW Definition 7,
  pp.~5179.} Intuitively, this variant simply makes the public good as
large as matching funds allow (i.e., large as to exhaust the matching
pool). In this paper, we first note that, because of the mentioned
matching requirements, this alternative mechanism is what will typically
be feasible in practice. It is also the case of what has been
implemented in Gitcoin Grants. In that platform, when the sum of
required payments to each project exceeds available matching funds
committed by donors, the subsidies to each project are scaled down by a
constant so totals add up to the subsidy pool's budget.\footnote{This is
  explained in Buterin (2019).} As a result (i.e., if contributors
perceive that matching funds will be scaled down to meet the funding
constraint) individual contributions are lower than the socially
optimal. Intuitively, this is because individuals are not fully
compensated by the social benefits they generate from contributing to
projects.

We next argue that, with limited matching funds, the relevant question
is if the mechanism is able to optimally allocate those \emph{limited
funds}. In Section \ref{laefficiency} we note that an optimal allocation
of a limited pool of matching funds should equalize the marginal social
benefits across projects. It turns out that this is not exactly the case
with the CQF allocation, since the resulting allocation entails
differences between the marginal benefits across projects. We show that
this deviations from optimality will be higher: i) the lesser funds are
available in the matching pool relative to matching requirements, ii)
the higher the variability in the supporting preferences across projects
(e.g., more equally invested projects imply higher marginal benefits
than more concentrated projects), and iii) the higher the number of
contributors.

Next, in terms of analyzing incentives to strategic behavior, in Section
\ref{backing}, we propose a simple formalization to analyze incentives
for strategic behavior. In particular, we analyze incentives facing
individuals raising funds in the mechanism (e.g., founders or team
members of public good related projects) to contribute to other projects
in the platform with the expectation of being invested back, a behavior
we call \emph{reciprocal backing}. We derive the conditions under which
such a strategy is profitable and propose related hypotheses.

In the second part of the paper, we explore evidence from Gitcoin Grants
7th and 8th rounds, which took place during 19 days between September
and October, 2020 and during 16 days in December 2020, respectively. In
Section \ref{gitcoin} we derive a series of stylized facts on individual
contribution to projects, and examine some of the proposed theoretical
hypotheses.

In Section \ref{kevolution}, we document that, consistently with the
theoretical intuition, QF target levels reached the funding constraint
quite rapidly, particularly in the main round categories. In these
categories this happened in less than four days, with a few reaching
this value just in the second day. We will note that this observation
has the implication that projects started to compete for funds very
early in their rounds. We also confirm the theoretical insight that
project matching fund requirements follow a quadratic relation with
respect to the number of contributors.

Second, in Section \ref{descriptives}, we show that projects' matching
requirements are somewhat aggravated by the fact that there is a
tendency among contributors to make very small investments, scattered
among multiple projects. This tendency can be seen, as we discuss, as
one of the empirical characteristics of this mechanism, different from
other documented crowdfunding mechanisms.

While such behavior is consistent with an interest of contributors in
promoting many projects -powered by QF matching-, it could also result,
as we mentioned, from strategic behavior. For example, we note that in
the specific case of reciprocal backing, this strategy provides positive
returns even when the pool of funds constraint is active. We document
that \emph{at least} 20\% of contributions in the rounds analyzed are
reciprocal, meaning that for every ten contributions to other projects
by project team members, two are invested back by invested
projects.\footnote{Due to possible limitations in the data source (which
  might not reflect all relationships between team members and
  projects), we understand that 20\% is a lower bound on the reciprocal
  investments that really take place in the platform.}

Finally, in Section \ref{regressions}, we specifically explore the
question of whether contributors internalize the matching budget
constraint. With this purpose, we estimate models that explain
contribution amounts as a function of the matching budget constraint. We
find a negative \emph{relationship}, which in principle is consistent
with the idea that individuals reduce their contributions as they
perceive their projects of interest will receive fewer matching funds.
We also explore an unexpected increase in the pool of matching funds
that occurred during the 7th round. In this case, we find that
contributors did not react by increasing their contributions. While this
might be rather contradictory in first sight, we show that this result
can be explained by the fact that although the pool of matching funds
increased by 25\%, due to the quadratic behavior of the mechanism, the
effect on the matching budget constraint was negligible.

We end the paper with a discussion that reviews its main insights and
suggests design implications.

\emph{Related literature}

This paper is related to the literature that studies quadratic financing
mechanisms for the optimal provision of public goods (Buterin, Hitzig,
and Weyl 2019). The quadratic mechanism as a form of collective action
pricing was proposed, and its equilibrium properties described, in Weyl
(2012) and Lalley and Weyl (2019). Other antecedents of mechanisms based
on quadratic mechanisms were proposed by Groves and Ledyard (1977) and
Hylland and Zeckhauser (1979). The literature on mechanisms for
nearly-optimal collective decision making is vast and one of its main
references is the Vickrey -- Clarke -- Groves preferences revelation
mechanism (Vickrey 1961; Clarke 1971; Groves 1973).

Arguably, QF can also be categorized as a form of \emph{crowdfunding}.
In crowdfunding, entrepreneurs raises external financing from a large
audience (the``crowd''), in which each individual provides a very small
amount (Belleflamme, Lambert, and Schwienbacher 2014). In exchange,
individuals pre-order a product or receive a -small- share of future
returns. While Gitcoin contributors do not receive shares of the
projects they invest in, they are arguably the beneficiaries of the
public goods (e.g., open source software, blockchain community growth,
etc.) they fund.\footnote{Note that Meyskens and Bird (2015) also
  includes donations as a form of crowdfunding.} As it discusses
characteristics of a form of crowdfunding the paper also aims to
contribute to that literature (See, for instance, (Short et al. 2017;
Agrawal, Catalini, and Goldfarb, n.d.; Burtch, Ghose, and Wattal 2015).

Finally, Gitcoin provides an example of an implementation of a new
financing mechanism for open sources projects. In that sense, the paper
adds, generally, to a vast literature that studies open source projects
from an economic and management perspective (von Krogh and von Hippel
2006; Nagle 2019; August, Chen, and Zhu 2021), and specifically to
financing issues related to these projects (Overney et al. 2020; Nakasai
et al. 2017)

\hypertarget{the-qf-rule-and-its-matching-fund-requirements}{%
\section{\texorpdfstring{The QF rule \label{QFrule} and its matching
fund
requirements}{The QF rule  and its matching fund requirements}}\label{the-qf-rule-and-its-matching-fund-requirements}}

The QF mechanism proposed by BHW is an allocation rule for the funding
of public goods. Funding comes from the support of individual
contributors plus an amount of matched funds provided by external
donors.\footnote{Importantly, recall that individual contributors are
  not required to fund the matching pool. As mentioned above, BHW show
  that introducing such a requirement changes the properties of the
  mechanism in terms of efficiency.} Under this mechanism, individual
contributions are inputs to determine which and how much individual
projects will be financed, while the external pool of matching funds
serves to match individual decisions (i.e., no allocation decisions are
made by external donors).

Assume that \(p\in P\) indexes public good projects competing to receive
funding. Also \(i \in I\) indexes individual contributors. An individual
\(i\) supports a project \(p\) by committing an amount of money
\(c_i^p\). We will examine the individual contributor decision problem
below. For the moment, assume individual contributions from the \(I\)
individuals to the \(P\) projects are known. In addition, assume there
is a pool of funds provided by donors, that we will denote \(D\), and
that will be used to match individual contributions. For the moment
assume that there is an endless pool of matching funds available (i.e.,
\(D\rightarrow+\infty\)).

\begin{figure}
\hypertarget{fig:qfrule}{%
\centering
\includegraphics[width=4.16667in,height=3.77083in]{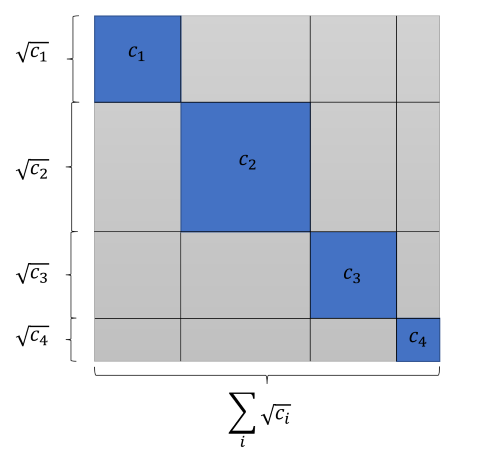}
\caption{QF rule. A public good with four contributions
\((c_1,c_2,c_3,c_4)\). The area of the outer square (i.e., sum of blue
and gray areas) represents the size of the public good according to the
QF rule \(F^{p,\text{QF}}=(\sum\sqrt{c_i})^2\). Total funds contributed
by individuals (i.e., \(C^p=\sum_i c_i^p\)) are represented by the sum
of the blue square areas. Total required matching funds (i.e.,
\(M^{p,QF}=F^{p,QF}-C^p\)) are represented by the gray
area.}\label{fig:qfrule}
}
\end{figure}

In such a context, the QF rule allocates, for each project \(p\), an
amount of funds we denote \(F^{p,\text{QF}}\), resulting from summing
the square roots of all individual contributions, and taking the square
of the result : \[
F^{p,\text{QF}}=\bigg(\sum_{i}\sqrt{c_i^p}\bigg)^2
\] Denote \(C^p\) as the total funds committed by individual
contributors to project \(p\), i.e., \(C^p=\sum_i c_i^p\). In order to
satisfy the QF rule, project \(p\) should additionally receive the
target matching amount \(M^{p,QF}\), defined by: \[
M^{p,QF}=F^{p,QF}-C^p
\] Figure \ref{fig:qfrule} presents a graphical representation of the QF
rule, first proposed in Buterin (2019). In the figure, total
contributions by individuals (\(C^p\)) are represented by the blue
shaded area. Required matching funds (\(M^{p,QF}\)) are represented by
the gray shaded area. Total funds proposed by the QF rule
(\(F^{p,\text{QF}}\)) are represented by size of the outer square (i.e.,
the sum of both blue and gray areas). The figure serves to illustrate
the size of matching fund requirements, as well as the fact that the
mechanism always requires positive external funds to work (i.e., there
is no case in which positive matching funds are not required).

\begin{observation} \label{thm:quadratic}

The total target matching amount $M^{p,QF}$ scales quadratically in the number of contributors.

\end{observation}

It is useful to notice that the level of LR subsidy a project \(p\)
receives can also be expressed as: \begin{equation}
M^{p,\text{QF}}=(\sum_i \sqrt{c_i^p})^2-\sum_i c_i^p= \\
\sum_i c_i^p+2\sum_{i\neq j}\sqrt{c_i^pc_j^p}-\sum_i c_i^p=2\sum_{i\neq j}\sqrt{c_i^pc_j^p}
\label{eq:qfmatch}\end{equation} This expression is useful because the
summation has a number of terms equal to the number of \emph{pairs} of
contributors. So while total individual contributions (\(C^p\)) scale
linearly, target matching amounts (\(M^{p,\text{QF}}\)) scale
quadratically following the number of pairs of contributors, which
corresponds to the combinatorial number \(n\choose2\)
=\(\frac{n!}{(n-2)!2!}=\frac{n (n-1)}{2}\approx \frac{n^2}{2}\).

In other words, the level of funding required to philanthropists needs
to scale as fast as the square of the number of contributors. Note that
this is certainly a demanding requirement (if not impossible) in any
application with a considerable number of contributors.

\begin{figure}
\hypertarget{fig:qf_matchs}{%
\centering
\includegraphics[width=4.16667in,height=3.77083in]{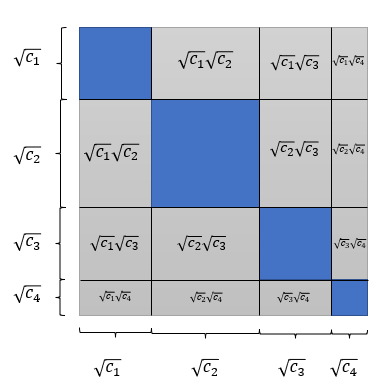}
\caption{Representation of the target matching amount according to the
QF rule. The size of the total matching amount, represented as the area
in gray, is comprised of rectangles of areas sized by
\(\sqrt{c_i^pc_j^p}\), which sum \(2\sum_{i\neq j}\sqrt{c_i^pc_j^p}\) as
in Equation \ref{eq:qfmatch}}\label{fig:qf_matchs}
}
\end{figure}

Notice that the last expression of Equation \ref{eq:qfmatch} can also be
identified using the same graphical representation. Figure
\ref{fig:qf_matchs} details that the target matching amount is comprised
of set of rectangles of areas sized by \(\sqrt{c_i^pc_j^p}\), and which
sum a total area of \(2\sum_{i\neq j}\sqrt{c_i^pc_j^p}\).

An alternative way of visualizing the requirements of the mechanism on
the matching fund is to consider the requirements of the marginal
contributor.

\begin{observation} \label{thm:quadratic}

A new contributor $j'$ contributing  $c_{j'}^p$ to project $p$ will require the QF mechanism an additional match of $\sqrt{c_k^p}\bigg(2\sum_{i\neq k}\sqrt{c_i^p}\bigg)$.

\end{observation}

In other words, an additional contributor places an increasingly
demanding burden on the matching fund, since it demands the mechanism to
match an amount that results from multiplying the new contribution by
all existing contributions to the mechanism summed.

\hypertarget{contribution-patterns-across-projects-and-required-matching-funds}{%
\subsubsection{Contribution patterns across projects and required
matching
funds}\label{contribution-patterns-across-projects-and-required-matching-funds}}

Finally, in inspecting the requirements on the matching fund it is also
useful to consider how different patterns of contributions across
projects pose different requirements to the matching fund.

Because of its quadratic design, given a fixed amount of total
contributions \(C^p\), the QF rule allocates a higher amount of matching
funds to projects with more contributors (with smaller contributions).
The flip side is that, from the point of view of fund requirements,
holding target QF constant, more equally (less concentrated) invested
projects require higher amounts of matching funding. This is expressed
in the following proposition:

\begin{proposition} \label{thm:funding}

Denote $\alpha_i=\frac{\sqrt{c_i}}{\sum\sqrt{c_i}}$ as the share of total of the square roots of funds contributed by individual $i$ , and $\sigma^2$ a measure of the variability among shares (i.e., $\sigma^2=\frac{1}{n}\sum_i(\alpha_i-\bar{\alpha})^2$, where $\bar{\alpha}=\frac{1}{n}\sum_i\alpha_i$). Then, total matching fund requirements are given by
$$
M^{p,\text{QF}}=(1-n\sigma^2-n\bar{\alpha}^2)F^{p,\text{QF}}
$$
As result, given a value of QF target, more equally invested projects require a higher amount of matching funds. 

\end{proposition}

The proof is in the Appendix.

As a simple illustration of this, Figure \ref{fig:reqfunds} presents two
projects, with a same QF target. Because Project 1 receives more equal
contributions than Project 2 , it also requires more matching funds.

\begin{figure}
\hypertarget{fig:reqfunds}{%
\centering
\includegraphics[width=6.25in,height=3.125in]{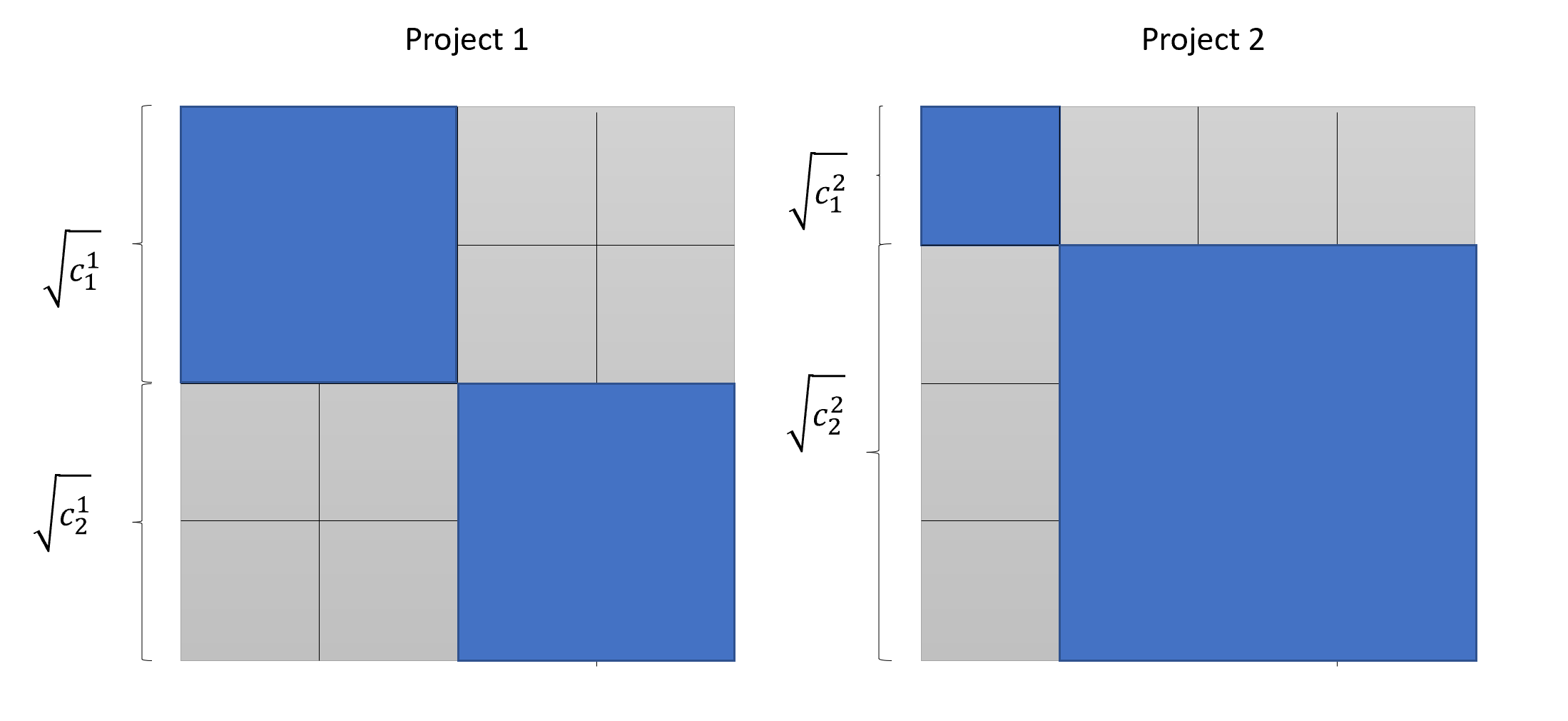}
\caption{Two projects with equal QF funding targets. More equal
contributions in Project 1 imply that this project requires a higher
amount of matching funds}\label{fig:reqfunds}
}
\end{figure}

This leads to the question of how different patterns of contributions
with individuals with given budgets will impact the matching
requirements. Precisely, assume there are \(N\) contributors, each with
a budget \(m_i\) to be invested across \(P\) projects. Denote \(s_i^p\)
the share of funds that the individual contributor will allocate to
project \(p\) , so \(\sum_{p}s_i^p=1\) \(\forall i\).

One observation in this case is that matching requirements will be
maximized if contributors' preferences, as measured by their share of
invested wealth, are perfectly correlated across individuals.

\begin{proposition} \label{thm:correlation}

Given $N$ contributors, each having a budget of $m_i$ and allocating $s_i^p$ to project $p$, matching fund requirements will be maximized if support preferences are perfectly correlated across individuals (i.e., for any two projects $p$ and $p'$ and any two individuals $i$ and $j$,  $\frac{s_i^p}{s_i^{p'}}=\frac{s_j^p}{s_j^{p'}}$)

\end{proposition}

See the proof in the Appendix.

This result has the corollary that the subsidy will be maximized under
complete coordination, for instance, if investments by all contributors
are allocated to a single project, if all contributors are coordinated
to invest half of the investments in two projects, and so on. In any of
these cases the total (maximum) amount of required funds will be given
by \begin{equation}
M^{\text{QF, MAX}}=2\sum_{i\neq j}\sqrt{m_i^pm_j^p}
\label{eq:maxmatch}\end{equation} Equation \ref{eq:maxmatch} resembles
Equation \ref{eq:qfmatch}, and retains the linearly scaling property in
terms of individuals wealth, and quadratically scaling in the number of
contributors.

In contrast, the total subsidy is minimized in case invested shares are
perfectly non-correlated, as in a case where each individual invests in
a separate project. Such a case would implies 0 matching funds.

\hypertarget{qf-with-limited-matching-funds}{%
\section{\texorpdfstring{QF with Limited Matching Funds
\label{QFwlimitedMF}}{QF with Limited Matching Funds }}\label{qf-with-limited-matching-funds}}

As we have discussed, in many cases, matching funds will be limited and
will not be enough to cover the QF target we discussed in the previous
section. Here we will not consider the case in which contributors are
required to cover deficits (i.e., differences between the QF matching
requirements and available funds). Instead, we will focus on the case in
which the mechanism is restricted to distribute the available pool of
matching funds.

Assume now that the matching pool has a size of \(D\) dollars. There are
\(P\) projects to match. Let \(M^{p}\) denote the amount of matching
funds that the mechanism will allocate to project \(p\) . The matching
funds constraint is: \[
\sum_pM^{p}\leq D
\] By the QF rule discussed above, assume each project has a target
match allocation of \(M^{p,\text{QF}}\). Assume that \(D\) cannot cover
allocating funds to all projects according to the QF rule. An additional
rule should added to the mechanism to distribute the limited matching
funds \(D\). We will take here the case of Gitcoin Grants, a platform we
will discuss in more detail in Section \ref{gitcoin}. Assume that, as in
the case of Gitcoin Grants, QF target allocations are scaled down by a
constant so the pool of matching funds constraint is satisfied in
equality\footnote{We will discuss the allocation mechanism of Gitcoin
  Grants in more detail in Section \ref{gitcoin}.}. Denote with \(k\)
such a constant, so \(k\) is chosen to satisfy: \begin{equation}
\frac{1}{k}\sum_pM^{p,\text{QF}}=D
\label{eq:kdefinition}\end{equation}

Once the balancing constant \(k\) has been defined, the matching
allocation rule is determined by: \begin{equation}
M^{p}=\frac{1}{k}M^{p,\text{QF}}=\frac{D}{\sum_p M^{p,\text{QF}}} M^{p,\text{QF}}
\label{eq:matchingrule1}\end{equation} Since projects also receive the
direct contributions \(C^p\), total funds that a project receives
(\(F^p\)) are given by: \begin{equation}
F^p=\frac{1}{k}M^{p,\text{QF}}+C^p=\frac{1}{k}(F^{p,\text{QF}}-C^p)+C^p=\frac{1}{k}F^{p,\text{QF}}+(1-\frac{1}{k})C^p
\label{eq:actualfunds}\end{equation}

In other words, the allocation rule is a mixture of QF with a weight on
unmatched private contributions. We note that this coincides with the
mechanism named by BHW as the capital-constrained quadratic finance
(CQF) mechanism\footnote{See See Buterin, Hitzig, and Weyl (2019),
  pp.~5179.}.

\begin{figure}
\hypertarget{fig:qf_limitedmatchs}{%
\centering
\includegraphics[width=6.25in,height=3.125in]{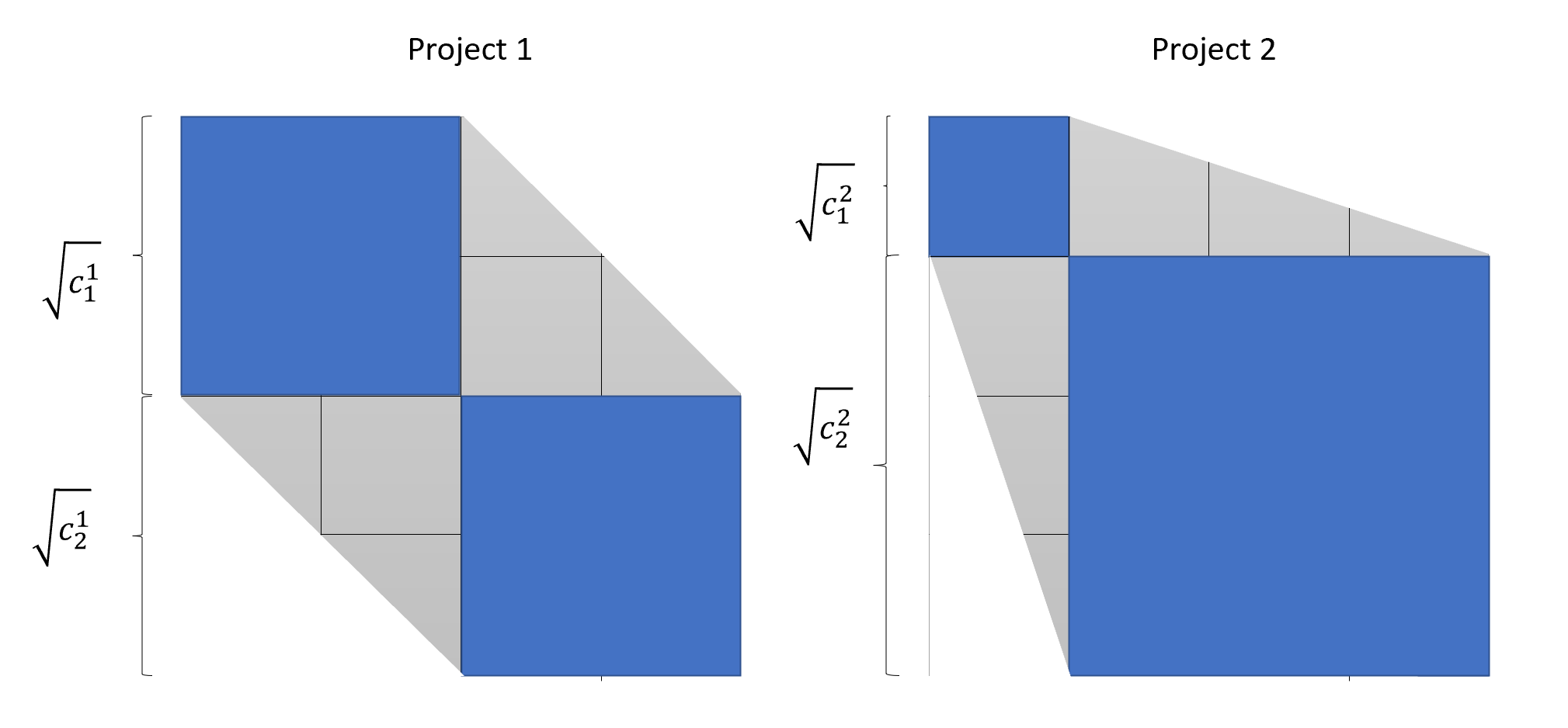}
\caption{Two projects with equal QF funding targets, and available
matching funds covering half of total required matching
funds}\label{fig:qf_limitedmatchs}
}
\end{figure}

To illustrate the resulting allocations, Figure
\ref{fig:qf_limitedmatchs} below represents an example with two projects
(1 and 2) and two contributors. In this example we assume contributions
are such that both projects determine an equal QF target funding (i.e.,
\(F^{1,\text{QF}}=F^{2,\text{QF}}\)). In addition, we assume that \(D\)
only covers half of target funds so the funds budget constraint is met
when \(k=2\). Gray areas in this case represent the effective matching
funds received by each project (i.e., \(M^{1},M^{2}\)) . Notice that in
this example \(M^{1}>M^{2}\), reflecting that, because of the QF rule,
more equal contributions demand more on the matching fund
(\(M^{1,QF}>M^{2,QF}\)). In this case, since target matching funds are
scaled down by a constant, all matching funds received by projects are
reduced proportionally.

\hypertarget{conditional-efficiency-of-the-decentralized-allocation}{%
\subsection{Conditional efficiency of the decentralized
allocation}\label{conditional-efficiency-of-the-decentralized-allocation}}

Here we revise some of the properties of the decentralized allocation
resulting from the mechanism. We start by reproducing BHW's result which
shows that the mechanism achieves a socially efficient allocation (i.e.,
an allocation that maximizes social welfare), in the case the pool of
funds is big enough to satisfy QF matching requirements. Since, as we
discussed, this is a demanding scenario, we explore the efficiency of
the allocation of \emph{limited funds}.

As in Buterin, Hitzig, and Weyl (2019), let \(V_i(F^p)\) be the
currency-equivalent utility a citizen \(i\) receives if the funding
level of public good \(p\) is \(F^p\). Utilities from different public
goods are assumed to be independent, and it is assumed a setting of
complete information.

\begin{observation}

The QF with limited matching mechanism tends to a generate a socially optimal allocation as matching funds tend to be enough to cover target requirements.

\end{observation}

The problem that defines the optimal individual contribution from the
perspective of backer \(i\) is\footnote{As in BHW, here we assume that
  \(k\) is unaffected by the individual decision \(c_i^p\). In addition,
  because of the independence of preferences among goods, we can examine
  the contribution to each project separately, and correspondingly, we
  drop the superscript \(p\) from the individual contribution \(c_i\) .}
\[
\max_{\{c_i\}}  V_i\bigg(\frac{1}{k}(\sum_{i}\sqrt{c_i})^2+(1-\frac{1}{k})C^p\bigg)-c_i
\] Where we have substituted Equation \ref{eq:actualfunds}\footnote{As
  mentioned above this is the Capital Constrained Quadratic Funding
  problem in Buterin, Hitzig, and Weyl (2019).}. This problem has a
First Order Condition (F.O.C.) given by: \begin{equation}
V_i'(F^p)(\frac{1}{k}\frac{\sum_{i}\sqrt{c_i}}{\sqrt{c_i}}+(1-\frac{1}{k}))=1
\label{eq:cpo}\end{equation}

Notice that if \(k \xrightarrow{} 1\) (i.e.,
\(\sum_p M^{p,\text{QF}}\xrightarrow{} D\)) then the F.O.C. condition
converges to: \begin{equation}
V_i'(F^p)=\frac{\sqrt{c_i}}{\sum_{i}\sqrt{c_i}}
\label{eq:FOCdecentralized}\end{equation} Summing Equation
\ref{eq:FOCdecentralized} across individuals gives the socially optimal
condition: \[
\sum_iV_i'(F^p)=1
\] In other words, the marginal cost of investing 1 unit of contribution
equals the aggregate marginal benefit for the community. This is the
standard efficient condition that a centralized planner would follow to
maximize aggregate welfare at cost \(c\). Indeed, in the case
\(k \xrightarrow{} 1\) the mechanism converges to the QF mechanism
discussed in Section \ref{QFrule}.

Also notice that if \(k \xrightarrow{} +\infty\) (i.e.,
\(\frac{\sum_p M^{p,\text{QF}}}{D}\xrightarrow{} +\infty\)) , the F.O.C
is\\
\[
V_i'(C^P)=1
\] Which is the socially inefficient, \emph{private} condition.

It follows that enough funds to guarantee the QF matching requirements
should be in place in order to sustain full efficiency.

Notice that if we additionally consider the requirements on the pool of
matching funds discussed in the previous section, the mechanism will not
obtain social efficiency in most practical applications.

\hypertarget{efficiency-in-terms-of-the-allocation-of-limited-funds}{%
\subsection{\texorpdfstring{Efficiency in terms of the allocation of
limited funds
\label{laefficiency}}{Efficiency in terms of the allocation of limited funds }}\label{efficiency-in-terms-of-the-allocation-of-limited-funds}}

A related question is into what extent the QF mechanism provides an
efficient solution to the problem of allocating a \emph{limited} pool of
matching funds. First, it is useful to recall what would be such
condition in the first place. A social planner with limited funds \(D\)
would maximize the aggregate welfare subject to the financing constraint
as follows: \[
\begin{split}
\max_{\{F^p\}^{p\in P}} \sum_i\sum_pV^{p}_i(F^{p}) \\\textrm{s.t.} \sum_p F^{p}=D
\end{split}
\] Notice that the F.O.C. for each project \(p\) are \[
\sum_iV'^{p}_i(F^{p})-\lambda=0 \;\;\;  \forall p
\] As result, an socially optimal allocation of limited funds would
equalize the sum of marginal utilities across projects.

\[
\sum_iV'^{p}_i(F^{p})=\sum_iV'^{p'}_i(F^{p'}),\ \forall p, p'
\]

To examine the extent in which the the CQF equalizes marginal benefits
across projects, we rearrange Equation \ref{eq:cpo} and sum across
individuals to obtain \begin{equation}
\sum_i V_i'(F_p)=\sum_i(\frac{1}{k}\frac{\sum_{i}\sqrt{c_i}}{\sqrt{c_i}}+(1-\frac{1}{k}))^{-1}  \;\;\;  \forall p
\label{eq:multiplier}\end{equation}

\begin{observation}

In general, the CQF mechanism does not equalize marginal social benefits of limited matching funds across projects.

\end{observation}

Define the RHS of Equation \ref{eq:multiplier} as \[
\lambda_p \equiv \sum_i(\frac{1}{k}\frac{\sum_{i}\sqrt{c_i}}{\sqrt{c_i}}+(1-\frac{1}{k}))^{-1}
\] Then, one dimension of the relative \emph{inefficiency} of the
mechanism allocation is given by the variability of this
\emph{multiplier} across projects. Notice, in addition, that this
magnitude is observable, and we will explore its empirical behavior in
Section \ref{kevolution}.

To understand the sums of the marginal valuations can vary between
projects, it is useful to note, first, that for given availability of
matching funds \(k\), and number of contributors (\(N\)), \(\lambda_p\)
will be higher for more equally invested projects.

\begin{proposition} \label{thm:lambdapandconcentration}

Given a value of matching requirements to available funds ($k$), and number of contributors ($N$), the multiplier $\lambda_p$ is lower bounded by 
$$
\lambda_p \geq\frac{n^2}{\frac{1}{k}\sum_i\frac{1}{\alpha_i^p}+n(1-\frac{1}{k})}
$$
which increases with more concentrated contributed projects. This implies that more equally contributed projects face a higher value of the multiplier $\lambda_p$.

\end{proposition}

The proof is in the Appendix.

Second, note that \(\lambda_p\) is an increasing and concave function
with respect to \(k\).\footnote{It is easy to verify that the second
  derivative of the function with respect to \(k\) is negative. We omit
  the proof for brevity.} The fact that the functional form is concave
makes the marginal penalty in terms of efficiency relatively higher for
low values of \(k\). This implies that the level of the relative
inefficiency will also be affected by this factor.

Finally, note that \(\lambda_p\xrightarrow{}N\) (i.e., the marginal
social cost under a private allocation) as \(k\) increases (i.e., as
fewer matching funds are available). In line with what was shown above,
this implies that as there are fewer matching funds, the mechanism will
converge to the (socially inefficient) private cost. It also implies
with more contributors (as \(N\) increases) there is a larger space for
differences in multipliers across projects .

\begin{figure}
\hypertarget{fig:ksimulation}{%
\centering
\includegraphics[width=3.85417in,height=2.60417in]{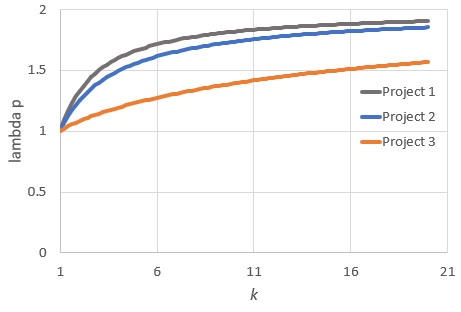}
\caption{Simulation of multiplier \(\lambda_p\) for three projects as
\(k\) increases. Each project have the same amount of target quadratic
funding (i.e., \(F^{1,\text{QF}}=F^{2,\text{QF}}=F^{3,\text{QF}}\), and
only two contributors. Project 1 is equally contributed by both
contributors. In Project 2 one contributor allocates twice of funds that
the other. In Project 3, one contributor allocates 15 times more funds
than the other.}\label{fig:ksimulation}
}
\end{figure}

To further illustrate how the sums of the marginal valuations can vary
between projects, consider Figure \ref{fig:ksimulation}, which
illustrates the behavior of \(\lambda_p\) for three projects that have
the same amount of target quadratic funding (i.e.,
\(F^{1,\text{QF}}=F^{2,\text{QF}}=F^{3,\text{QF}}\)), and only two
contributors (\(N=2\)). Contributions for Projects 1 and 2 correspond to
those illustrated in Figure \ref{fig:qf_limitedmatchs} (i.e., In Project
1, both contributors contribute the same amount, and in Project 2, one
contributor contributes twice as much as the second.) In Project 3 one
of the contributors contributes 15 times the contribution of the second.
The figure shows how \(\lambda_p\) evolves as \(k\) increases for each
project (i.e., less available matching funds in relation to target
required funds). The figure illustrates how more equally invested
projects tend to converge to the inefficient marginal cost faster in
terms of relative funding availability.

We can conclude that the allocation of a limited pool of matching funds
deviates more from a social optimal allocation: i) the lesser the amount
of funds in the pool relatively to total QF target matching fund
requirements, ii) the greater the variability in the contribution
patterns across projects (i.e., variability in terms of how concentrated
or equally invested the projects are), and iii) the higher the number of
contributors.

\hypertarget{collusion-and-reciprocal-backing}{%
\section{\texorpdfstring{Collusion and reciprocal backing
\label{backing}}{Collusion and reciprocal backing }}\label{collusion-and-reciprocal-backing}}

QF has \emph{collusion} and \emph{identity fraud} as central
vulnerabilities (\emph{BHW}). While fraud refers to the idea of a
participant misrepresenting itself as many (a.k.a., Sybil attack),
collusion might take various forms, including agreements between
contributors or with other agents outside the mechanism. In this section
we focus on problems of the latter type, particularly in incentives for
strategic behavior.

Consider two contributors that have candidate projects in the mechanism,
and that decide to invest in each other. We might call such a situation
``reciprocal backing''. This behavior is interesting because it's a form
of \emph{reciprocity}, that has been observed in many settings (Fehr and
Gächter 2000; Göbel, Vogel, and Weber 2013). We note that such situation
could arise from an explicit collusive agreement, or from an implicit
behavior. To analyze the economic incentives, we will analyze the cases
in which contributors could increase their own payoffs by investing a
share of their funds in each other instead of fully backing their own
projects. Assume that each contributor has an amount \(c\) to invest,
and there are no limits for matching funds. Consider the strategy of
investing half of the funds in the project of another contributor with
the expectation of being invested back. In such a situation the
strategic decision takes the form of 2 by 2 simultaneous game, with the
following net payoff matrix.

\begin{table}[!htbp]

\centering

\begin{tabular}{lcc}

 & \multicolumn{1}{l}{Invest} & \multicolumn{1}{l}{Do not invest} \\ \cline{2-3} 

\multicolumn{1}{l|}{Invest} & \multicolumn{1}{c|}{$c$,$c$} & \multicolumn{1}{c|}{$-\frac{c}{2}$,$c\frac{(1+2\sqrt{2})}{2}$} \\ \cline{2-3} 

\multicolumn{1}{l|}{Do not invest} & \multicolumn{1}{c|}{$c\frac{(1+2\sqrt{2})}{2}$,$-\frac{c}{2}$} & \multicolumn{1}{c|}{$0$,$0$} \\ \cline{2-3} 

\end{tabular}

\end{table}

Where, for instance, the upper-left \emph{Invest-Invest} payoff is given
by

\(\big(2\sqrt{\frac{c}{2}}\big)^2-c=c\).\footnote{If both follow
  ``Invest'', then the payoff for individual 1 is the quadratic funding
  rule for two investments sized \(\frac{c}{2}\) minus the cost \(c\),
  therefore \(\big(2\sqrt\frac{c}{2}\big)^2-c=c\). If both follow ``Do
  not invest'', then they just receive the quadratic rule for their
  individual investments of \(c\), therefore \((\sqrt{c})^2=c\). If
  Individual 1 invests but Individual 2 does not invest back then
  Individual's 1 payoff is the quadratic funding rule of just half of
  her investment \(\big(\sqrt\frac{c}{2}\big)^2-c=-\frac{c}{2}\). The
  outcome of Individual 2 in that case is
  \(\big(\sqrt\frac{c}{2}+\sqrt{c}\big)^2-c=c(\frac{1+2\sqrt{2}}{2})\)}

The resulting game is a standard \emph{Prisoners' Dilemma}, with a
non-collusion Nash equilibrium
\(\text{("Do not invest", "Do not invest")}\).

This example illustrates, as pointed out by BHW, that the collusion
problem is mitigated because of unilateral incentives to deviate from
the collusive agreement.\footnote{See Buterin, Hitzig, and Weyl (2019),
  pp.~5180.}

A first point to note, however, is that in the practice of QF, rounds
might take a repetitive form. For instance, in \emph{Gitcoin Grants},
projects can participate in every round that take place every two or
three months. By December 2020, eight rounds were already closed, and
Gitcoin planned to continue organizing rounds, since its aim is to
provide a sustained flow of financing for projects. If the strategic
game presented above is played infinite times, or if there is
uncertainty when it will stop, a collusion can be sustained as a Nash
equilibrium, using \emph{trigger strategies}, or threads (Friedman
1971). This leads to the following proposition:

\begin{proposition}\label{thm:collusioninrepetitivegames} 

Incentives for strategic behavior, taking the form of multilateral reciprocal contributions, are part of a Nash equilibrium when players can participate in a indefinite number of rounds.

\end{proposition}

See the proof in the Appendix.

When the number of participants in this type of collusion increases
above two, the collusion strategy is still profitable when a percentage
of the participants deviate. In the no-funding limits case, for example,
a collusion strategy with \(n\) participants is still profitable under
deviation if a percentage of \(\alpha^*\)\footnote{Note that if there is
  no restriction on matching funds, if a percentage of \(\alpha\) of the
  \(n\) contributors invests in the reciprocal strategy, then the
  project receives an amount, given by the quadratic rule, of
  \((\alpha n \sqrt{\frac{c}{n}})^2\). For a contributor such strategy
  is profitable if \((\alpha^{*} n \sqrt{\frac{c}{n}})^2-c>0\). Then the
  required percentage of contributors participating in the reciprocal
  strategy is \(\alpha^{*}> \sqrt{\frac{1}{n}}\). Following Proposition
  \ref{thm:collusioninrepetitivegames} it follows that this strategy can
  be sustained in an equilibrium with an indefinite number of rounds.}
still colludes, where \begin{equation}
\alpha^*>\frac{1}{\sqrt{n}}
\label{eq:alpha}\end{equation}

So, for instance, a reciprocal contribution strategy with 25 investments
is profitable if 20\% of invests back.

A more realistic scenario is when there is a limited pool of matching
funds. Under restrictive funds (\(k>1\)), collusion incentives are
reduced as the pool of funds dry up (BHW). But note this does not happen
as fast as one could expect. The following proposition illustrates this
observation:

\begin{proposition} \label{thmcollusionn}

Under restrictive funds a collusion strategy with $n$ participants is profitable under deviation if a percentage of $\alpha^{**}$ still colludes where

\end{proposition}

\begin{equation}
\alpha^{**}>\frac{k\bigg((1-\frac{1}{k})+\sqrt{(1-\frac{1}{k})^2+4(\frac{n}{k})}\bigg)}{2n}
\label{eq:eqalphadstar}\end{equation}

\begin{figure}
\hypertarget{fig:alpha}{%
\centering
\includegraphics[width=5.125in,height=2.98958in]{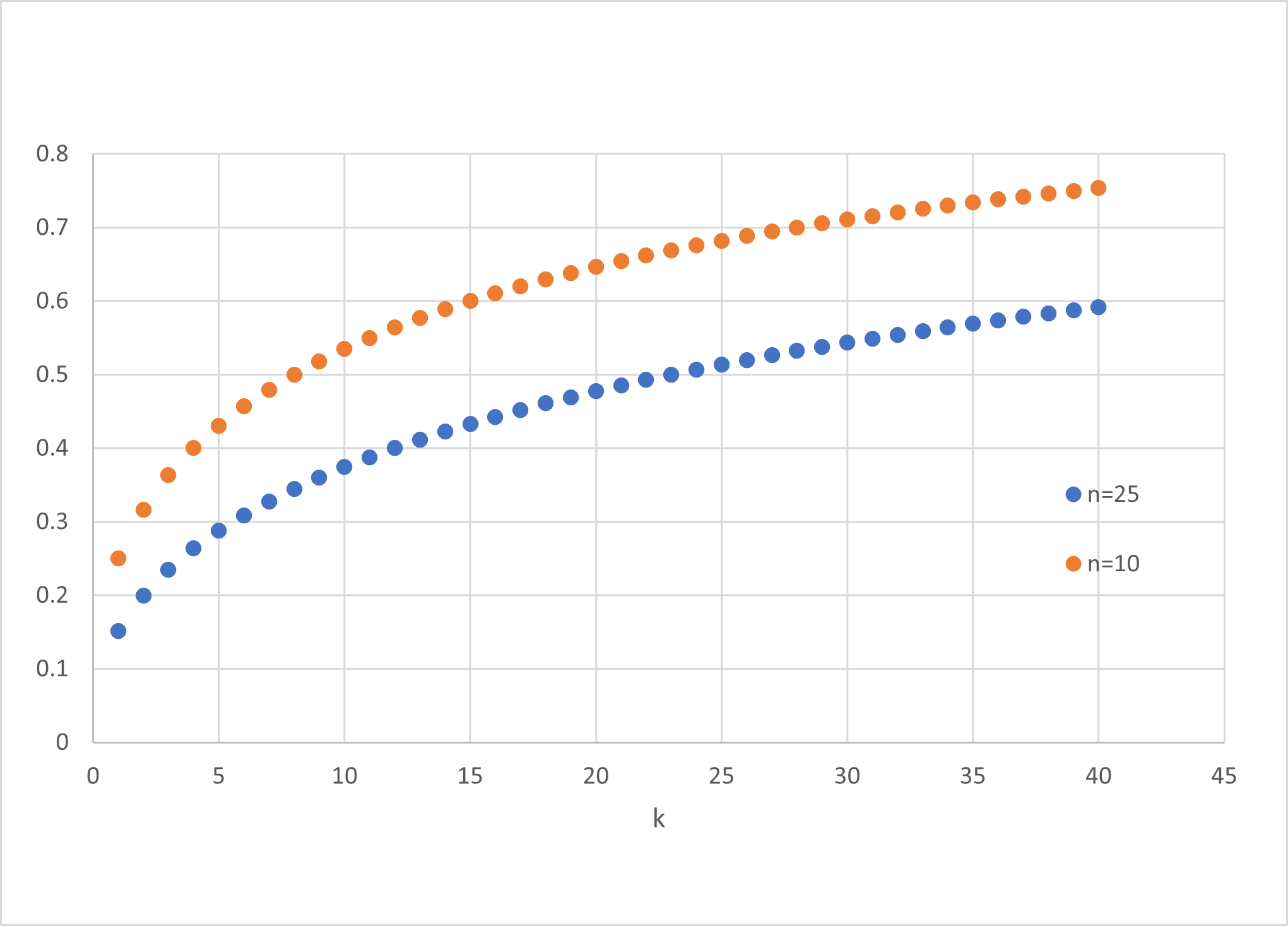}
\caption{Minimum percentage or participants non deviating for collusion
to be profitable}\label{fig:alpha}
}
\end{figure}

Following the previous example, a reciprocal investment strategy with 25
investments and, for example, (k=20), is profitable if \emph{60\%} of
contributors invests back. As we will confirm in the empirical section
below, values of (k) between 10 a 20 are in line with what is taking
place in Gitcoin rounds. Figure \ref{fig:alpha} illustrates further the
result in Equation \ref{eq:eqalphadstar}. It depicts two cases of
collusions sized \(n=10\) and \(n=25\) respectively, and shows the
required percentage of participants for the reciprocal strategy to be
profitable, against different levels of (k). The figure serves to
illustrate that there are feasible percentages of participation rates in
reciprocal strategies that can serve to sustain an equilibrium.

A final point to note is that, in reciprocal contributions taking place
in practice, not all participants are subject to the same budget
constraint. As we will note in the case of Gitcoin, because rounds for
different \emph{categories} take place simultaneously, each with its own
budget constraint, colluding projects do not compete for funding if they
are part of different categories. This simple observation leads to the
following proposition:

\begin{observation}

\label{thmreciprocalacross}

Incentives for strategic behavior in the form of reciprocal investments are higher across project categories than within categories. 

\end{observation}

\hypertarget{qf-evidence-from-gitcoin-grants}{%
\section{\texorpdfstring{QF evidence from Gitcoin Grants
\label{gitcoin}}{QF evidence from Gitcoin Grants }}\label{qf-evidence-from-gitcoin-grants}}

\hypertarget{details-of-gitcoin-grants-7th-and-8th-rounds}{%
\subsection{Details of Gitcoin Grants' 7th and 8th
rounds}\label{details-of-gitcoin-grants-7th-and-8th-rounds}}

The 7th Gitcoin Grants round took place between September 15th and
October 2nd, 2020. The 8th round took place between the 2nd and 17th of
December 2020. The 7th round was organized around three main categories:
Infra Tech, Applications (\emph{DApps}) Tech, and Community Projects. A
fourth specific category, \emph{Matic Network} (technology
infrastructure used for scalability) started at the same time. The first
three categories received initially a matching endowment of 120 thousand
DAI (as we will explain below, on September 23th, the pool of funds
increased to 150 thousand DAI each). The \emph{Matic} endowment received
50 thousand DAI.

The 8th Gitcoin Grants Round presented the same main three categories
(DApps, Infra and Community), each endowed with 100 thousand DAI, and
three additional categories: Filecoin Liftoff (projects related to
Filecoin, the decentralized storage network) with 100 thousand DAI,
Apollo (also an initiative related to Filecoin) with 50 thousand DAI,
and East-Asia (an initiative to support projects from East-Asia) with 50
thousand DAI.

\begin{figure}[H]

\centering

\includegraphics[scale=0.2]{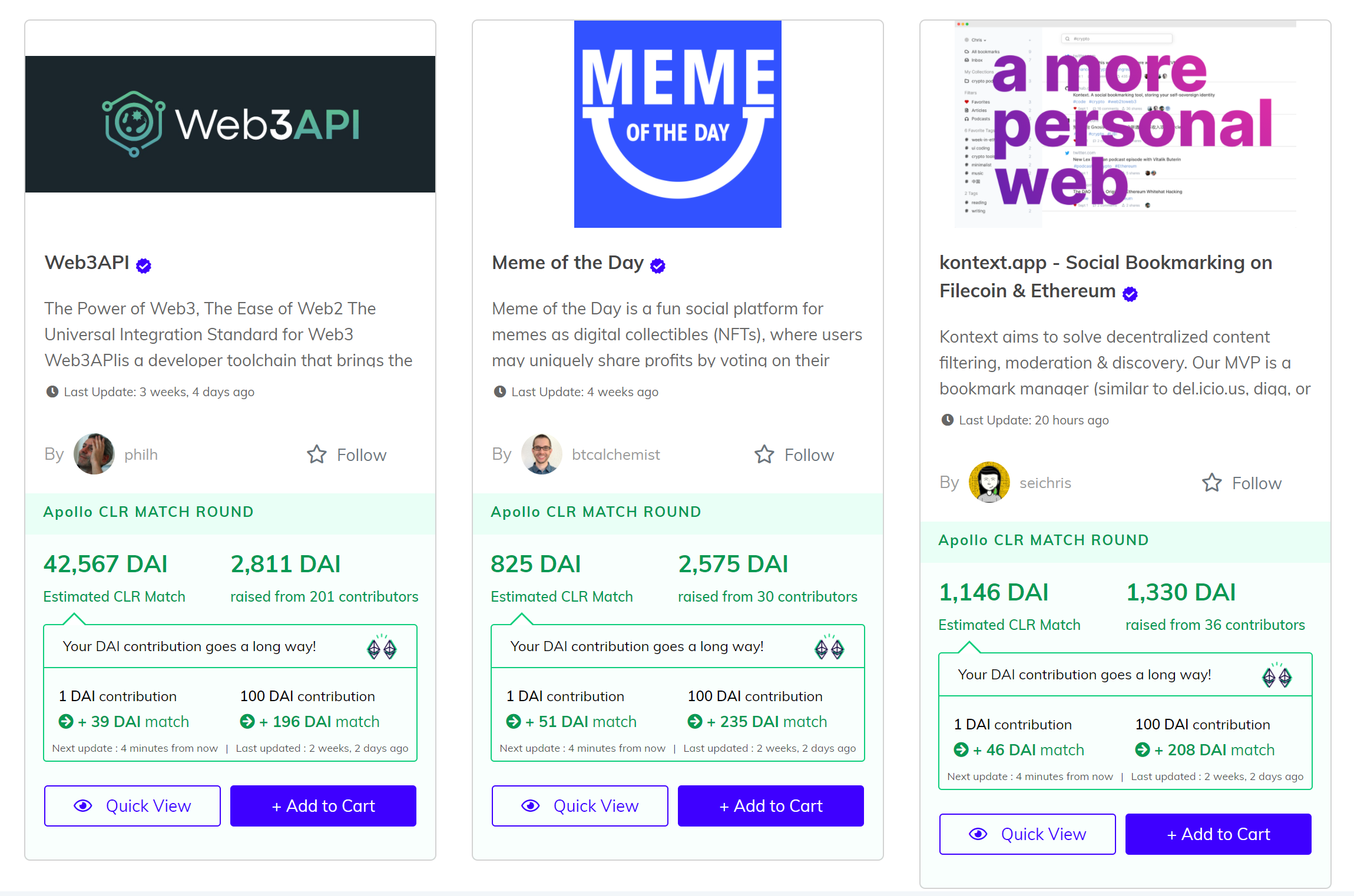}

\caption{Gitcoin Grants webpage extract}\label{fig:extract}

\end{figure}

During the days of a round, contributors could easily find the
participating projects by browsing Gitcoin Grants' webpage, choose which
projects to invest in and commit a cryptocurrency transfer. The page
reports the total amount received so far by each project, and
importantly, provides an estimation of the expected amount to be
received by the project in terms of matching funds if an individual
contributes (See Figure \ref{fig:extract}). For this, the website
continuously computes the specific allocations taking into account the
budget constraint, in line with the discussion above.

\hypertarget{projects-contributors-and-amounts}{%
\subsection{\texorpdfstring{Projects, contributors and amounts
\label{descriptives}}{Projects, contributors and amounts }}\label{projects-contributors-and-amounts}}

Table \ref{tbl:sumstats1} displays descriptive statistics on the number
of projects, contributors, and total individual contributed amounts per
project and category. A total of 249 projects and 1,234 contributors
participated in Round 7. The category with the highest number of
contributors was \emph{DApps Technology} with 93 projects (37\% of total
projects) and 1105 contributors (90\% of the contributors). The second
most important category in terms of contributors was \emph{Community}
with 82 projects (33\%) and 677 contributors (54\%). \emph{Infra tech}
displayed less projects (52, 33\%), but a similar number of contributors
(613, 49\%). Finally, the smallest category was \emph{Matic}, with 19
projects and 190 contributors. Notice that the sum of contributors
percentages exceeding 100\%, shows that contributors tend to invest in
many projects. Round 8 counted with more projects (444) and contributors
(4,953), and presented a similar pattern in terms of importance of the
main categories (\emph{DApps}, Community and \emph{Infra}). Statistics
for Round 8 are available in Table \ref{tbl:sumstats8} in the Appendix.

\begin{center}

\begin{table}[!htbp]

\caption{Gitcoin Round 7 Descriptive Statistics\label{tbl:sumstats1}}

\small

\begin{tabular}{p{6cm} p{1.5cm} p{1.5cm} p{1.5cm} p{1.5cm} p{1.5cm}}
\toprule
 & N & Mean & Std. dev. & Std. error & Median \\
\midrule
\textit{All categories. projects: 249, contributors: 1234} &  &  &  & \\
$c_i$ & 8450 & 24.895 & 161.778 & 1.76 & 4.75 \\
$\sqrt{c_i}$ & 8450 & 2.829 & 4.112 & 0.04 & 2.179 \\
\textit{Dapp Tech category. projects: 93, contributors: 1105} & \textit{} & \textit{} &  &  \\
$c_i$ & 3802 & 19.407 & 120.596 & 1.956 & 3.64 \\
$\sqrt{c_i}$ & 3798 & 2.466 & 3.654 & 0.059 & 1.908\\
\textit{Infra Tech category. projects: 56, contributors: 613} & \textit{} & \textit{} &  &  \\
$c_i$ & 2220 & 39.301 & 247.737 & 5.258 & 4.75\\
$\sqrt{c_i}$ & 2220 & 3.370 & 5.287 & 0.112 & 2.179\\
\textit{Community category. projects: 82, contributors: 677} & \textit{} & \textit{} &  &  \\
$c_i$ & 2142 & 21.366 & 115.481 & 2.495 & 4.75\\
$\sqrt{c_i}$ & 2142 & 2.958 & 3.553 & 0.077 & 2.179\\
\textit{Matic category. projects: 19, contributors: 190} & \textit{} & \textit{} &  &  \\
$c_i$ & 345 & 10.613 & 39.070 & 2.103 & 3.3\\
$\sqrt{c_i}$ & 345 & 2.401 & 2.205 & 0.119 & 1.817\\
\bottomrule
\end{tabular}

\vspace{1ex}

\begin{tabular}{cc}

\multicolumn{2}{l}{Note: This table reports summary statistics on backer contributions per project.}

\end{tabular}

\end{table}

\end{center}

Individual contributions per project \((c_i^p\)) average about 25 DAI
(Table \ref{tbl:sumstats1}) in Round 7 and about 30 DAI in Round 8, but
these figures can be quite misleading given the fact that the
distributions of contributions are severely right-skewed, with 80\% of
contributions below 10 DAI, 17\% between 10 and 100 DAI, and only 3\%
above 100 DAI in Round 7.

Histograms in Figures \ref{fig:histround7} (for Round 7),
\ref{fig:histcip_round8a} and \ref{fig:histcip_round8b} (for Round 8) in
the Appendix reflect this fact. The figures display histograms of total
individual contributions per projects by categories, where the range was
split in 3 intervals to more accurately visualize the respective
frequencies. The graphs show that for all categories contributions are
concentrated in values less than 10 DAI.

This pattern of small contributions are notably lower than what has been
documented in other crowdfunding settings. For instance, according to
Mollick (2014), contributions per backer in \emph{Kickstarter} for
technology projects averaged 73 USD.

\begin{figure}
\hypertarget{fig:histround7}{%
\centering
\includegraphics[width=5.58333in,height=7.55208in]{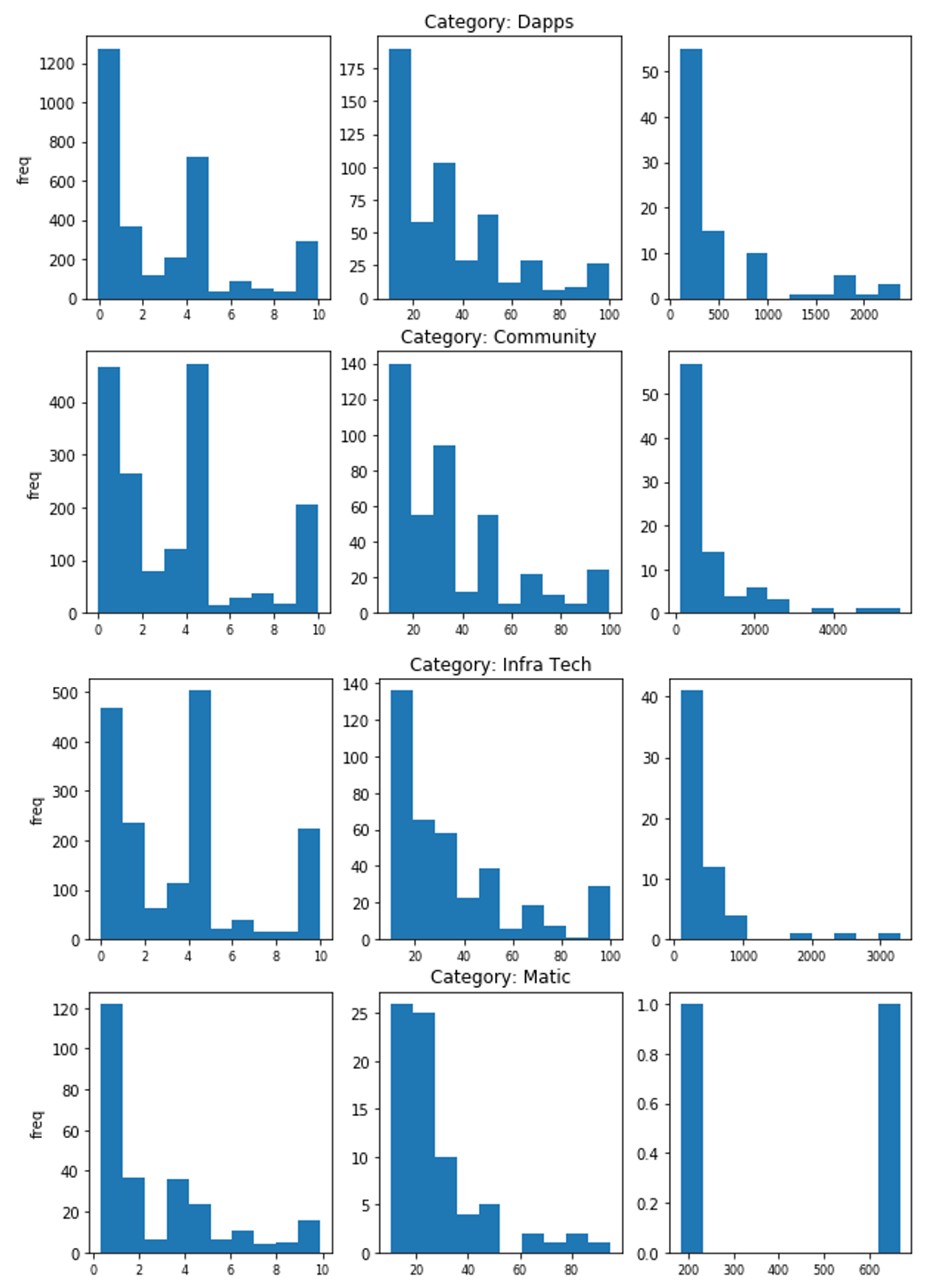}
\caption{Split range histograms of total individual contributions to
projects \(c_i^p\) by category. Round 7}\label{fig:histround7}
}
\end{figure}

\emph{Total contributions per individual and projects invested}

\begin{figure}
\hypertarget{fig:histmi}{%
\centering
\includegraphics[width=5.02083in,height=6.94792in]{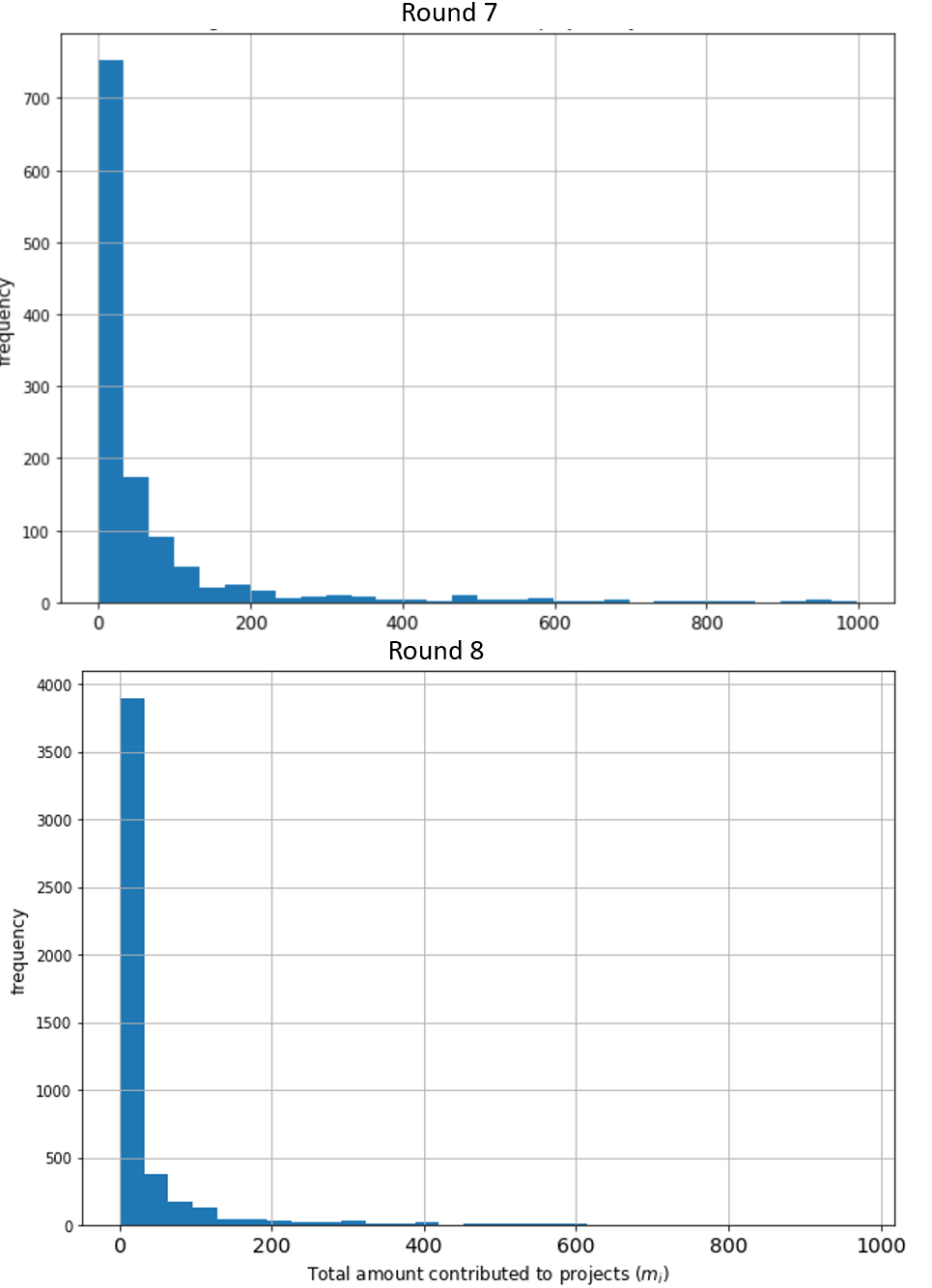}
\caption{Histograms for the total amount contributed to projects by
individuals (\(m_i\)). Rounds 7 and 8}\label{fig:histmi}
}
\end{figure}

In terms of the total contributions per backer (a proxy of \(m_i\)) the
median was 20.57 DAI and the mean 170.55 DAI also presenting an
asymmetric distribution (Figure \ref{fig:histmi}). In the case of Round
7, for example, nearly 32\% of individuals contributed less than a total
of 10 DAI, 50\% contributed between 10 and 100 DAI, 15\% of individuals
contributed between 100 and 1,000 DAI, and only 2\% contributed above
1,000 DAI.

\begin{figure}
\hypertarget{fig:histnp}{%
\centering
\includegraphics[width=5.02083in,height=6.94792in]{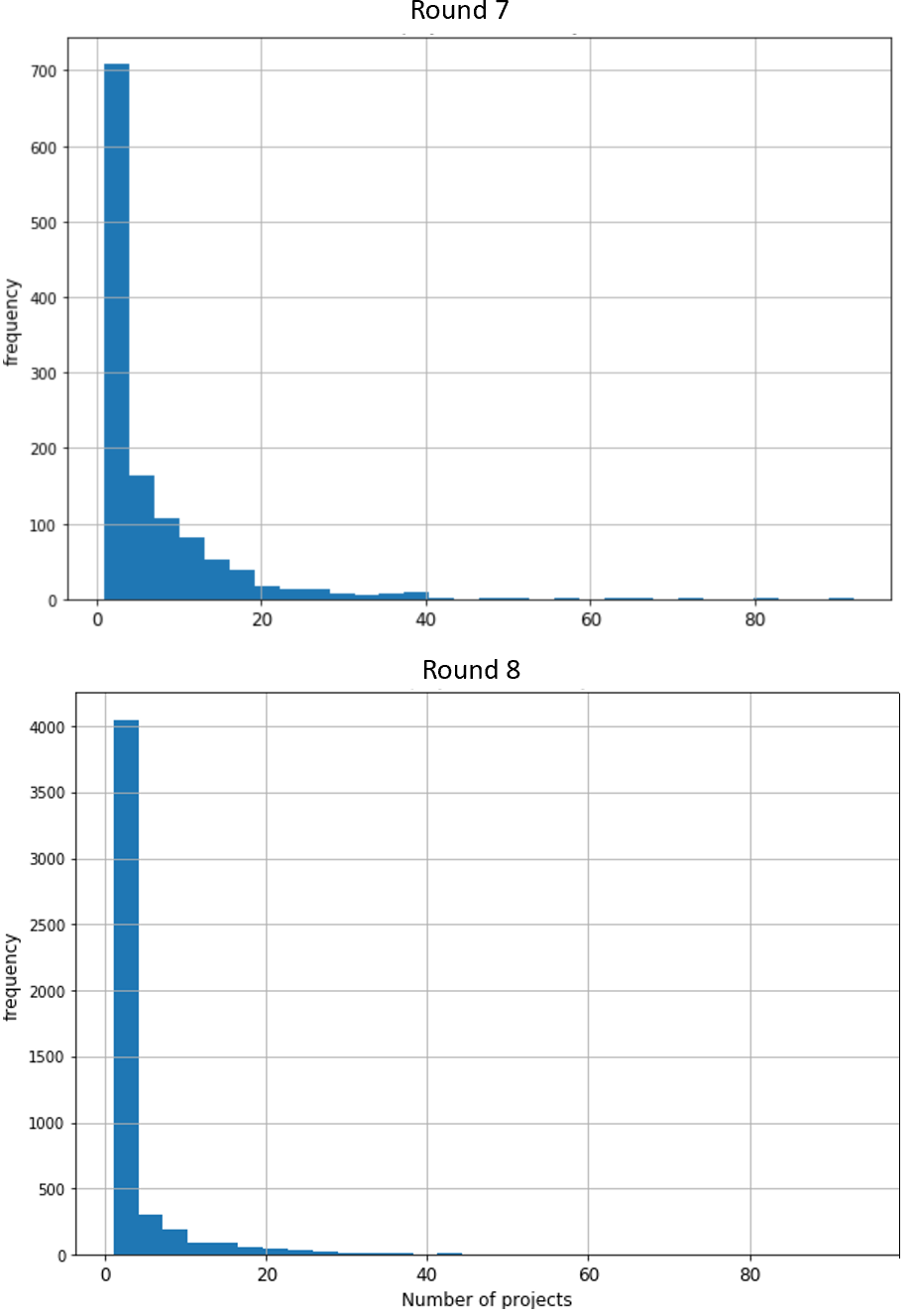}
\caption{Histograms for the number of projects contributed by individual
contributors. Rounds 7 and 8}\label{fig:histnp}
}
\end{figure}

This pattern of small investments is also consistent with the average
number of contributions per backer climbing to 6.85 on average, and a
median of 3 contributions. This is also a highly asymmetric distribution
as shown in Figure \ref{fig:histnp} which shows there is an economic
significant number of contributors investing in many projects, with 20\%
investing above 10 projects, and 6\% investing above 20.

\hypertarget{evolution-of-matching-fund-requirements-to-available-matching-funds-k-projects-deficits-and-measured-efficiency}{%
\subsection{\texorpdfstring{Evolution of matching fund requirements to
available matching funds (\(k\)), projects deficits and measured
efficiency
\label{kevolution}}{Evolution of matching fund requirements to available matching funds (k), projects deficits and measured efficiency }}\label{evolution-of-matching-fund-requirements-to-available-matching-funds-k-projects-deficits-and-measured-efficiency}}

\begin{figure}
\hypertarget{fig:ktrend}{%
\centering
\includegraphics[width=4.47917in,height=4.375in]{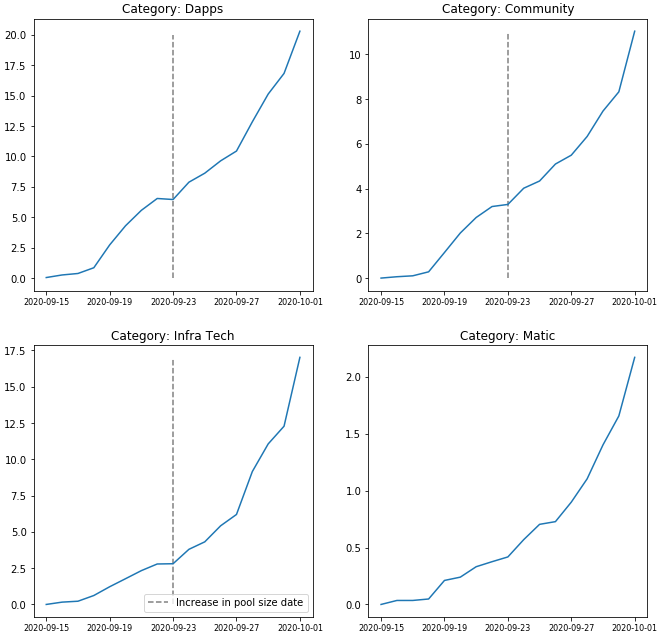}
\caption{Evolution of matching fund requirements to available matching
funds ratio (\(k\))}\label{fig:ktrend}
}
\end{figure}

The evolution of the constant \(k\) (See Equation \ref{eq:kdefinition})
during the round reflects how quickly required matching funds escalate.
Figure \ref{fig:ktrend} plots the constraint for each of the categories
in Round 7. Figures \ref{fig:round8_k_part1} and
\ref{fig:round8_k_part2} (in the Appendix) provide the plots for
categories in Round 8. These figures show that \(k\) increased non
linearly in all categories, reaching, for example, a value close to 20
for \emph{DApps} and \emph{Infra Tech} categories, and close to 12 for
the \emph{Community} category in Round 7. In Round 8, \(k\) reached even
higher values, approaching 50 for \emph{DApps} and \emph{Infra}, and 120
for the \emph{Community} category. This is not surprising given the fact
that the pool of funds for these categories was smaller in Round 8, and
the number of individuals contributing nearly quadrupled.

A different behavior resulted in the specific categories. Although these
categories do also show a rapid nonlinear increase in the value of
\(k\), because of the lower number of supporters, they ended with lower
\(k\) values. For instance, \emph{Matic} in the 7th round ended with a
value of \(k\) close to 2, \emph{Apollo} in round 8 ended with \(k\)
about 4, and \emph{Liftoff} ended with a value close to 0.2.

For graphs in Figure \ref{fig:ktrend}, the vertical dotted line in the
graphs indicates the day the pool of matching funds increased -recall as
explained above total funding increased in 25\%-. We can see that the
effect on \(k\) is a slight drop, which quickly resumes growth due to
the increase in the quadratic fund target. We will return to this point
in the next section.

\begin{figure}
\hypertarget{fig:deficits}{%
\centering
\includegraphics[width=5.73958in,height=2.1875in]{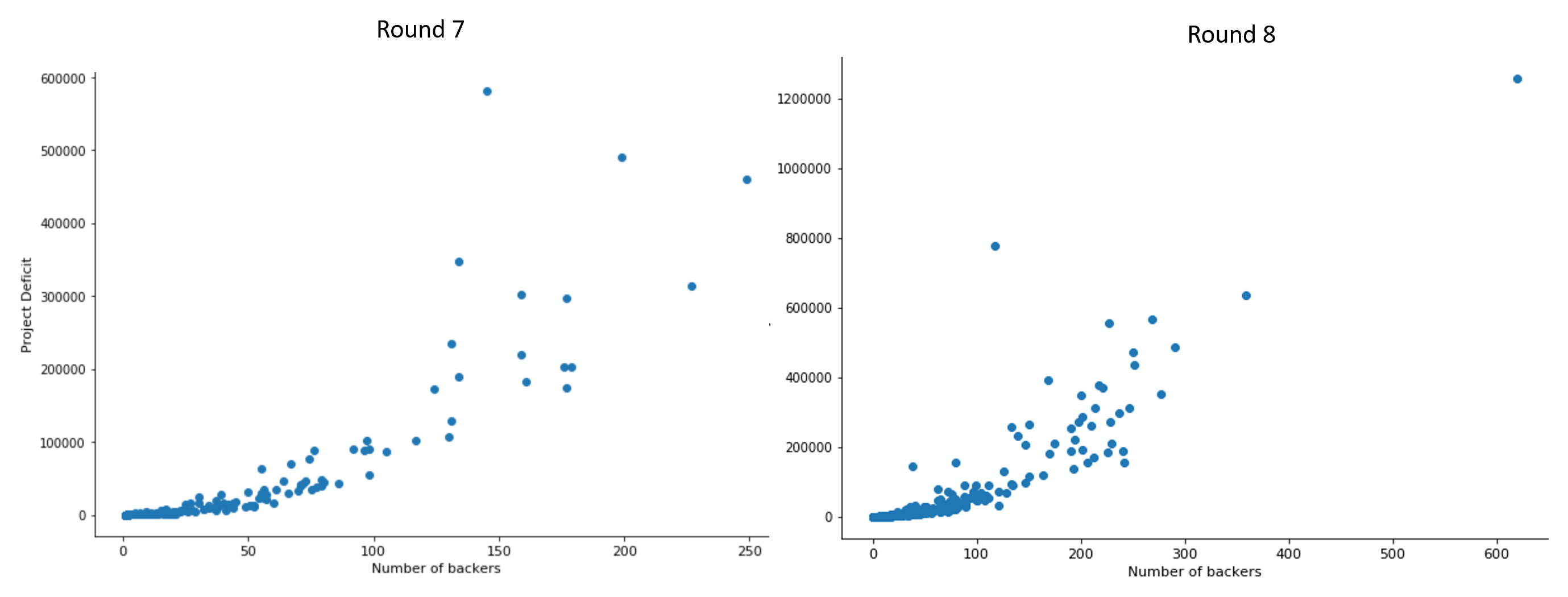}
\caption{Matching requirements and number of contributors at the end of
the round. Rounds 7 and 8}\label{fig:deficits}
}
\end{figure}

In terms of the determinants of \(k\) , Figure \ref{fig:deficits} is
also illustrative, showing the total matching requirements per project
versus the number of contributors. We can identify the relationship
proposed by Equation \ref{eq:qfmatch} above, where the behavior
increases quadratically. This relationship is confirmed in both rounds.

Finally, Table \ref{tbl:lambdap_ronda7y8} presents summary statistics on
\(\hat{\lambda_p}\). According to Equation \ref{eq:multiplier}, a
condition for the efficient allocation of limited funds is that the sum
of marginal benefits equalizes across projects. Therefore, the standard
deviation of \(\hat{\lambda_p}\) in Table \ref{tbl:lambdap_ronda7y8}
provide a measure of relative inefficiency in the allocation.

Table \ref{tbl:lambdap_ronda7y8} provides evidence that confirms the
theoretical predictions. In first place, categories with relatively more
funds available (i.e., a lower final \(k\) value), such the specific
categories Matic -Round 7-, East Asia and Liftoff -Round 8-, are the
categories with less variability in \(\lambda_p\) across projects. All
categories in round 8 resulted in higher variability than those in round
7. We also confirm that the higher the number of participants, the
higher the variability in \(\hat{\lambda_p}\), and lower efficiency in
the Round.

\begin{center}

\begin{table}[!htbp]

\centering

\caption{$\hat{\lambda_p}$ allocation efficiency indicator \label{tbl:lambdap_ronda7y8}}

\small

\begin{tabular}{p{3.5cm} p{1.8cm} p{1.8cm} p{1.8cm}}
\toprule
  Category &  Projects &       Mean &  Std. Dev. \\
\midrule
      Dapp - Round 7&      57 &  10.6060 &   4.6709 \\
 Community - Round 7 &      53 &   5.8891 &   2.6123 \\
     Infra - Round 7&      36 &  10.1602 &   3.2982 \\
     Matic - Round 7&      19 &   1.9273 &   0.1055 \\
 Community - Round 8 &     133 &  25.1127 &  21.2072 \\
      Dapp - Round 8&     114 &  16.6023 &  11.7582 \\
     Infra - Round 8&      43 &  22.8302 &  10.3061 \\
 East-Asia - Round 8&      23 &  14.7344 &   5.3482 \\
   Liftoff - Round 8&       3 &   0.2325 &   0.0793 \\

\bottomrule
\end{tabular}

\vspace{1ex}


\end{table}

\end{center}

\hypertarget{reciprocal-backing}{%
\subsection{\texorpdfstring{Reciprocal backing
\label{reciprocal-backing}}{Reciprocal backing }}\label{reciprocal-backing}}

\begin{figure}
\hypertarget{fig:reciprocal}{%
\centering
\includegraphics[width=6.44792in,height=2.54167in]{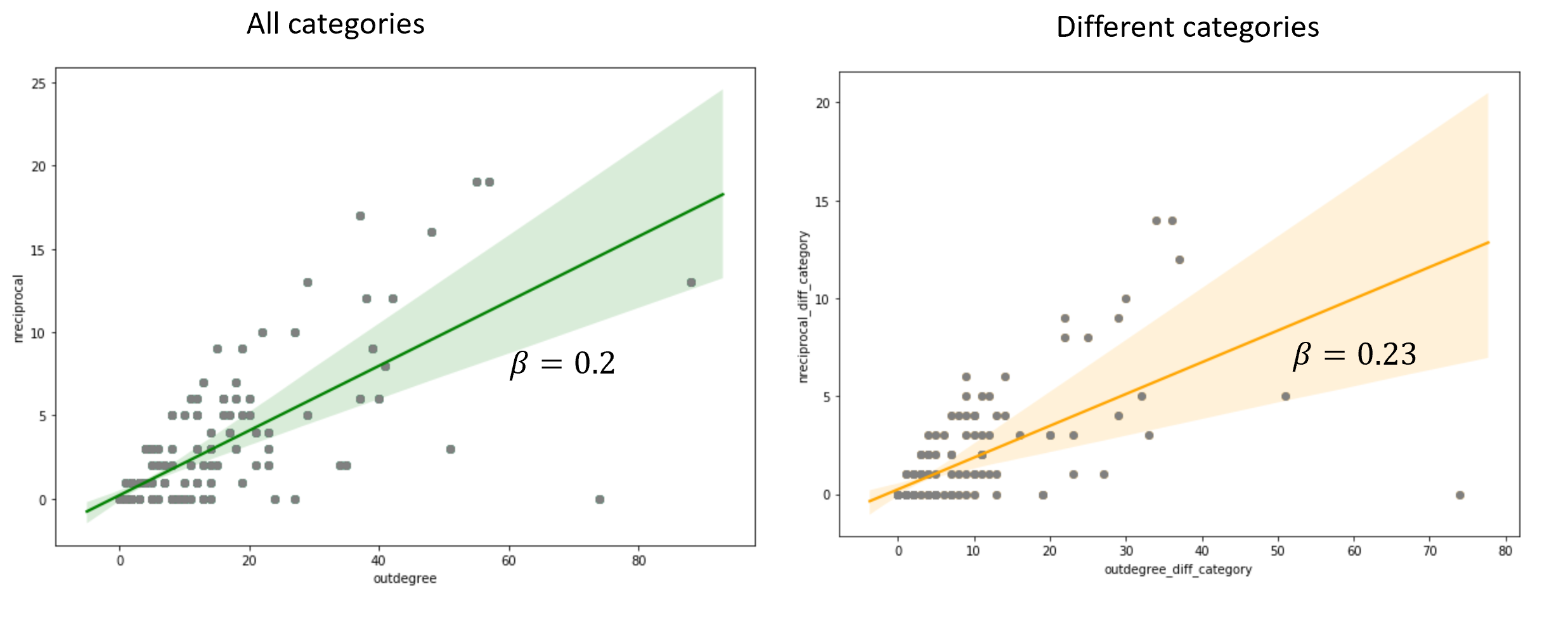}
\caption{Reciprocal contributions as function of total
contributions}\label{fig:reciprocal}
}
\end{figure}

In Section \ref{backing}, we discussed that under certain conditions, QF
provides incentives for strategic contributions in the form of
reciprocal backing. Figure \ref{fig:reciprocal} and Table
\ref{tbl:reciprocal_percentages} provide evidence on the extent of such
behavior.

To measure \emph{recyprocal backing} we exploit information on the
identity of projects' team members as registered in Gitcoin. Precisely,
we define contributions as \emph{reciprocal} if team members of a
project receiving contributions, support back the projects of their
contributors. Figure \ref{fig:reciprocal} illustrates the measure. On
the left panel, the figure plots each project number of reciprocal
contributions against the total number of contributed projects (i.e.,
total outdegree in networks terminology). The total number of
contributed projects is similarly constructed, by aggregating all
contributions by the project's team members. A linear approximation to
this relationship retrieves a slope of 0.2, suggesting that
approximately 20\% of the contributions are received back. The measure
depicted in the right panel is further restricted to contributions
across projects that belong to \emph{different} categories. Proposition
\ref{thmreciprocalacross} above, argued that reciprocal backing
incentives were stronger across categories. Here the figure suggests
that a contribution in an additional project is related with a
probability of reciprocity of 23\%. So, while the data confirms the
cross-category hypothesis, it also suggests that the magnitude of these
incentives are low.

We can further examine reciprocal investments across categories in Table
\ref{tbl:reciprocal_percentages}. Here we consider reciprocal
contributions taking into account the relative importance of the
respective categories. Column I displays the number of projects for each
category as percentage of the total number of projects in the Round.
Column II displays the complement of this percentage (i.e.~projects that
are not in the same category). If contributions were made with
independence of the project category, the percentage of reciprocal
investments across categories would tend to equal the percentage of
projects in Column II. In other words, if reciprocal backing incentives
were greater across categories, the resulting percentages would be
greater than those in Column II. We confirm this hypothesis in only one
of the main three
categories.\footnote{We omit considering the specific categories because of their relative size}
Projects in the \emph{Community} category, tend to have 71\% of their
reciprocal contributions in other categories, while only 66\% of
projects are not in the \emph{Community} category. This behavior,
however, is reverted in the \emph{DApp} category (55\% of reciprocal
contributions in other categories against 66\% of projects in other
categories), and no significant differences appear in the
\emph{Infra Tech} category.

Overall, we conclude that reciprocal backing is present in the data but
to a small extent (not explaining more than 20\% of investments), and
that incentives for reciprocal investments across categories are also
small.

\begin{center}

\begin{table}[!htbp]

\centering

\caption{Reciprocal backing.\label{tbl:reciprocal_percentages}}

\small

\begin{tabular}{p{3.5cm} p{2.5cm} p{2.5cm} p{2.5cm}}
\toprule
   Category &  Number of projects \% &  Number of projects- complement \% &  Reciprocal investments in other categories \% \\
\midrule
  Community &              33.49 &                         66.51 &                                      71.30 \\
  Dapp Tech &              33.49 &                         66.51 &                                      55.55 \\
 Infra Tech &              22.64 &                         77.36 &                                      76.00 \\
      Matic &              8.96 &                         91.04 &                                      20.51 \\
\bottomrule
\end{tabular}

\end{table}

\end{center}

\hypertarget{individual-contributions-and-availability-of-matching-funds}{%
\subsection{\texorpdfstring{Individual contributions and availability of
matching funds
\label{regressions}}{Individual contributions and availability of matching funds }}\label{individual-contributions-and-availability-of-matching-funds}}

Limited availability of matching funds, as explained in Section
\ref{QFwlimitedMF}, implies that contributors will reduce their
contributions as they learn there are less matching funds available. In
this section we examine the relationship between contributions and the
availability of matching funds (as measured by the ratio \(k\)).

First, we will assume that individuals do not anticipate the level of
\(k\) at the closure of the round and adjust their contributions as
information on \(k\) is updated (i.e., as the round
progresses).\footnote{As displayed in Figure \ref{fig:extract}, Gitcoin
  Grants backers can find an estimation of the amount of value that will
  be paired by the matching fund as part of the information available
  for each project.} As required matching funds increase continuously
during the round, we hypothesize that contributions will decrease
steadily as \(k\) increases.

Second, we will additionally examine an event where the pool of matching
funds (exogenously) changed during the round. As a general case, at the
beginning of a round, Gitcoin Grants announces the total pool of
matching funds for the entire round. However, on one occasion (due to
the unexpected entry of philanthropic contributors), on September 23th,
2020, Gitcoin Grants announced an (unexpected) increase in the size of
the pool as the round unfolded {[}\^{}reason{]}. In this particular
event, total matching funds were increased by 25\% (from 120 to 150
thousand DAI) for the three main categories (\emph{Dapps},
\emph{Infrastructure} and \emph{Community}). The forth contemporaneous
category (\emph{Matic}) remained unchanged.

In general, an increase in matching funds is expected to stimulate
increases in backers' contributions. However, as noted in Section
\ref{kevolution}, once the round has progressed, requirements on funds
to be matched might be already high so that increases in the pool of
funds could result in insignificant changes in the corresponding value
of \(k\). As a result, even considerable increases in the pool of
matching funds could result in no incentives to increase backers
contributions.

\hypertarget{econometric-specifications}{%
\subsubsection{Econometric
specifications}\label{econometric-specifications}}

To analyze the relationship between contributions and the level of \(k\)
during the round, we propose a simple econometric model of the form: \[
\sqrt{c_{i,p,t}}=\beta_0+\beta_1k_{t,\kappa}+\delta_p+\epsilon_{i,p}
\] Where \(k_{t,\kappa}\) is the constant balancing constraint as
defined in Equation \ref{eq:kdefinition}. The subscript \(t\)
acknowledges that \(k\) changes in time (as the round progresses), and
\(\kappa\) recognizes that \(k\) is specific to the project category
(See description on Gitcoin Rounds in Section
\ref{qf-evidence-from-gitcoin-grants}). \(\delta_p\) adds a project
specific fixed effect, and is introduced in some of the specifications
estimated.

Second, to examine the event where the pool of matching funds increased
during the 7th round, we propose a model of the form: \begin{equation}
\sqrt{c_{i,p,t}}=\beta_0+\beta_1t+\beta_2Post_t+\beta_3Increase_{\kappa}+\beta_4Post_t*Increase_{\kappa}+\delta_p+\epsilon_{i,p}
\label{eq:econometricmodel2}\end{equation} where \(t\) stands for the
number of the day in which the round took place, \(Post_t\) is a dummy
that is active on September 23th and after that date, and
\(Increase_{\kappa}\) is a dummy that is active for the categories that
increased funds at the mentioned date (i.e., active for \emph{Dapps},
\emph{Infrastructure} and \emph{Community}, and remains inactive for
\emph{Matic}).

Following the hypothesis that contributors will update the information
on \(k\) as round progresses, we expect \(\beta_1\) to capture the
negative effect on contributions. We also expect the coefficient on the
interaction \(Post_t*Increased\) (\(\beta_4\)) to identify a
differential effect (if any) on contributions after September 23th on
the round categories that experienced the matching funds pool increase.

\hypertarget{econometric-results}{%
\subsubsection{Econometric Results}\label{econometric-results}}

Table \ref{tbl:regressionsk} presents the results of the econometric
model of individual contributions, where the constant \(k\) is included
as an explanatory factor. In these specifications we consider the square
root level of the contribution as the dependent variable.

\begin{center}

\begin{table}[!htbp]

\centering

\caption{Effects of $k$ on contributions. Round 7 \label{tbl:regressionsk}}

\small

\begin{tabular}{lp{2.5cm} p{2cm} p{2cm} p{2cm}}
\toprule
    Dependent: Sqrt(c) &         (1) &        (2) &         (3) \\
\midrule
             Intercept &   2.9098*** &  1.9893*** &   2.0299*** \\
                       &    (0.0885) &   (0.3983) &    (0.3907) \\
         Category Dapp &      0.1086 &   1.0488** &   1.1628*** \\
                       &    (0.1027) &   (0.4136) &    (0.4145) \\
   Category Infra Tech &   0.4987*** &  1.2687*** &    1.0134** \\
                       &    (0.1235) &   (0.4484) &    (0.4507) \\
        Category Matic &  -0.6979*** &   0.8435** &   1.1431*** \\
                       &    (0.1205) &   (0.4226) &    (0.4313) \\
                     $k$ &    -0.0186* &  -0.031*** &  -0.0678*** \\
                       &    (0.0097) &   (0.0104) &    (0.0235) \\
       $k$*Category Dapp &             &            &       0.023 \\
                       &             &            &    (0.0263) \\
$k$*Category Infra Tech &             &            &    0.0737** \\
                       &             &            &    (0.0329) \\
      $k$*Category Matic &             &            &  -0.4104*** \\
                       &             &            &     (0.147) \\
  Project Fixed Effect &          No &        Yes &         Yes \\
        Number of obs. &        8650 &       8650 &        8650 \\
                Adj-R2 &       0.004 &      0.031 &       0.033 \\
          F-statistic: &      19.005 &    515.592 &     8049.32 \\
    Prob (F-statistic) &           0 &          0 &           0 \\
\bottomrule
\end{tabular}

\vspace{1ex}

\begin{tabular}{cc}

\multicolumn{2}{l}{Note: Project level clustered robust standard errors in parenthesis.}

\end{tabular}

\end{table}

\end{center}

Column 1 shows that the relationship with \(k\) is negative and
statistically significant, and the coefficient remains negative and
significant as project fixed effects are included. Column 3 additionally
allows the \(k\) effect to vary across categories. The results allow
calculating that the net effect of \(k\) was negative and significant in
the \emph{Community} (baseline, -0.0678), \emph{DApps}
(-0.0678+0.023=-0.04), and \emph{Matic} (-0.0678-0.4104=-0.4782)
categories. The exception is \emph{Infrastructure} category, where the
resulting effect is statistically not different from zero
(-0.0678+0.0737=0.0059).

Overall these results seem to confirm the hypothesis that backers update
their contributions as they learn on the state of the matching funds
constrain (i.e., that new contributions will be matched with lower
amounts by the matching fund).

Table \ref{tbl:regressionsDD1} and \ref{tbl:regressionsDD2} present the
results on the regressions that explore the consequences of the increase
in funds in the main categories in September 23th. Each table presents
variations on the baseline specification in Equation
\ref{eq:econometricmodel2}. Table \ref{tbl:regressionsDD1} includes
project fixed effects in Columns 3 and 4, and category-specific linear
trends in Columns 2 and 4. Table \ref{tbl:regressionsDD2} presents
results from reducing the sample of analysis to the immediate dates to
the event. Precisely, the sample is restricted to 2 days before and 2
days after the increase (Columns 1 and 2), and to 1 day before and 1 day
after the increase (Columns 3 and 4).

Table \ref{tbl:regressionsDD1} shows a negative and statistically
significant coefficient on \(t\), confirming a negative trend in
contributions as the round progresses. The coefficients on the
interaction \(Post_t*Increased\), however, show that there are no
significant changes associated with the increase in funds event.
Consistently, Table \ref{tbl:regressionsDD2} shows no sign of effects in
the immediacy of the event. \(Post\) tends to show negative but
insignificant changes. The coefficient on the interaction
\(Post_t*Increased\) is generally low in terms of its standard error and
therefore remains statistically insignificant across specifications.

\begin{center}

\begin{table}[!htbp]

\centering

\caption{Examining the pool of matching funds increase event. \label{tbl:regressionsDD1}}

\small

\begin{tabular}{p{3.5cm} p{1.8cm} p{1.8cm} p{1.8cm} p{1.8cm} }
\toprule
Dependent: Squared root of contribution &         (1) &         (2) &         (3) &         (4) \\
\midrule
                              Intercept &   1.9569*** &   2.0231*** &   1.8298*** &   1.8745*** \\
                                        &    (0.1388) &    (0.1497) &    (0.0907) &    (0.1133) \\
                          Category Dapp &      0.0223 &       0.037 &    1.083*** &    1.212*** \\
                                        &    (0.1347) &    (0.3231) &    (0.0748) &    (0.1966) \\
                    Category Infra Tech &    0.4735** &     -0.0332 &   1.3742*** &   0.9291*** \\
                                        &    (0.1885) &    (0.3815) &    (0.0706) &    (0.2561) \\
                         Category Matic &   0.7828*** &   0.7808*** &   1.5012*** &   1.5337*** \\
                                        &    (0.2274) &    (0.2716) &    (0.1579) &    (0.2228) \\
                                      t &  -0.0502*** &  -0.0641*** &  -0.0685*** &  -0.0843*** \\
                                        &     (0.018) &    (0.0221) &    (0.0177) &    (0.0222) \\
                                   Post &     -0.0703 &      0.0772 &       0.096 &       0.233 \\
                                        &    (0.4375) &    (0.3605) &    (0.4621) &    (0.3197) \\
                                Increased &    1.174*** &   1.2422*** &   0.3286*** &   0.3409*** \\
                                        &    (0.1376) &    (0.1972) &    (0.0749) &    (0.1205) \\
                                     Post*Increased &       0.329 &      0.1774 &      0.1923 &      0.0518 \\
                                        &    (0.4161) &    (0.3854) &    (0.4501) &    (0.3427) \\
                   Project Fixed Effect &          No &          No &         Yes &         Yes \\
                 Category linear trends &          No &         Yes &          No &         Yes \\
                         Number of obs. &        8650 &        8650 &        8650 &        8650 \\
                                 Adj-R2 &       0.005 &       0.006 &       0.033 &       0.033 \\
                           F-statistic: &     367.027 &     269.205 &      83.347 &      51.652 \\
                     Prob (F-statistic) &           0 &           0 &           0 &           0 \\
\bottomrule
\end{tabular}

\vspace{1ex}

\begin{tabular}{cc}

\multicolumn{2}{l}{Note: Project level clustered robust standard errors in parenthesis.}

\end{tabular}

\end{table}

\end{center}

\begin{center}

\begin{table}[!htbp]

\centering

\caption{Examining the pool of matching funds increase event. \label{tbl:regressionsDD2}}

\small

\begin{tabular}{p{3.5cm} p{1.8cm} p{1.8cm} p{1.8cm} p{1.8cm} }
\toprule
Dependent: Squared root of contribution &        (1) &        (2) &        (3) &        (4) \\
\midrule
                              Intercept &  1.6811*** &  1.7579*** &  1.2939*** &   1.942*** \\
                                        &   (0.1205) &   (0.1773) &   (0.1168) &      (0.2) \\
                          Category Dapp &     0.0471 &     0.1197 &  1.6631*** &    1.14*** \\
                                        &   (0.2032) &   (0.2991) &    (0.077) &   (0.1285) \\
                    Category Infra Tech &   0.5425** &     0.1294 &   2.074*** &  0.8225*** \\
                                        &   (0.2628) &   (0.3194) &   (0.0573) &    (0.071) \\
                         Category Matic &  0.6343*** &   0.6205** &  1.3915*** &  2.2013*** \\
                                        &   (0.2101) &   (0.3005) &   (0.2031) &   (0.4045) \\
                                   Post &    -0.2649 &  -0.9059** &     0.1015 &    -1.0707 \\
                                        &   (0.5075) &   (0.4078) &    (0.622) &   (0.7055) \\
                                Increased &  1.0468*** &  1.1374*** &    -0.0975 &    -0.2593 \\
                                        &   (0.1553) &   (0.2587) &   (0.1299) &   (0.2349) \\
                                     Post*Increased &     0.0499 &     0.6756 &    -0.3232 &    1.3205* \\
                                        &    (0.514) &   (0.4872) &   (0.6202) &   (0.7242) \\
                   Project Fixed Effect &         No &         No &        Yes &        Yes \\
                            Data window &     2 days &     2 days &      1 day &      1 day \\
                         Number of obs. &       1856 &        856 &       1856 &        856 \\
                                 Adj-R2 &      0.005 &     -0.001 &      0.124 &      0.332 \\
                           F-statistic: &      201.9 &    160.929 &     100.18 &    290.313 \\
                     Prob (F-statistic) &          0 &          0 &          0 &          0 \\
\bottomrule
\end{tabular}

\vspace{1ex}

\begin{tabular}{cc}

\multicolumn{2}{l}{Note: Project level clustered robust standard errors in parenthesis.}

\end{tabular}

\end{table}

\end{center}

\hypertarget{discussion-and-conclusions}{%
\section{Discussion and conclusions}\label{discussion-and-conclusions}}

Buterin, Hitzig, and Weyl (2019) (BHW) presents an innovative financing
mechanism for public goods with some very promising features. Our
interest in this paper has been to explore the matching fund
requirements of the mechanism and its implications in terms of
efficiency. In practice, as we have exemplified with the case of Gitcoin
Grants, the implementation of the QF mechanism will most likely take
place in the form of its \emph{capital-constrained} version (CQF). This
is because matching funds requirements increase fast, quadratically in
the number of contributors (Proposition \ref{thm:quadratic}). The
evidence also shows that there is a tendency among contributors to make
small contributions to multiple projects (Section \ref{descriptives}),
which also increase matching requirements. The tighter the restriction
on matching funds, the lower the social efficiency in the BWH sense.

Seeking to increase the matching pool of funds as a response seems
difficult in line with the evidence emerging from Gitcoin Grants. The
data illustrates that the funding restriction is reached fast, in the
first days of the rounds (Section \ref{kevolution}). Increases in
philanthropic funds might only provide a temporary relief, since funds
would eventually need to increase 20 to 50 times their size in some
cases -as can be observed in the value of \(k\) in Section
\ref{kevolution}-. An illustrative example is an event of funds
increasing 25\% in the middle of Round 7, which had almost negligible
effect on the ratio of required funds (to available funds) \(k\)
(Figures \ref{fig:ktrend} and \ref{fig:round8_k_part1}).

As a result, it is expected that projects will compete for the limited
pool of matching funds. It follows that a more appropriate question to
whether there will be some degree of inefficiency -relative to an
otherwise limitless funding scenario-, is the question of to what extent
the allocation of \emph{limited funds} is efficient. Social benefits
from available projects should equalize on the margin for such an
efficient allocation. (Section \ref{laefficiency}). It turns out that
that the CQF allocation entails some deviations from such an allocation.
Deviations are expected to be greater the lesser matching funds are
available, the higher the differences in patterns of contributions
across projects, and the higher the number of contributors. The data
from Gitcoin Grants also provides evidence in this respect. We have
shown that the mechanism did a better job in equalizing these benefits
when there were relatively more matching funds available (Section
\ref{kevolution}).

Other observations emerging from Gitcoin data seem particularly relevant
in terms of the implementation of the mechanism. For instance, the fact
that a higher correlation in contributions to projects among individuals
is related to higher needs of matching funds (Proposition
\ref{thm:correlation}) implies that features that ease or foster
contributions correlations among individuals have implications in terms
of funds requirements and efficiency. An example of such feature was
implemented during the 7th Gitcoin round. \emph{Grants Collections}
allowed any user to replicate a curated portfolio of contributions from
another contributor. While it is beyond the scope of this paper to
document the effect of the introduction of such a feature, it is worth
noting that such changes have effects that are worthwhile study.

Another important characteristic that emerges from the data analysis of
contributions is related to their small relative size (Section
\ref{descriptives}) (i.e., relative to those contributed in other
crowdfunding platforms such as Kickstarter). As mentioned above, small
contributions accelerates the needs for matching funds. While such
behavior could be attributable to the potential of quadratic backing per
se (while there are available matching funds funds), we have also noted
that this behavior might also be the result of strategic incentives. In
particular, contributors with listed projects might seek a return from
reciprocity by other contributors (Section \ref{backing}).

QF has certainly powerful properties, and we expect that there will much
more research on how to best implement QF ahead. This research should
take into account the economics of fund requirements, such as to how to
best allocate limited funds. This will be particularly important in
terms of implementing QF in applications associated with large
communities. How to discourage strategic behavior is another line of
further research.

\hypertarget{references}{%
\section*{References}\label{references}}
\addcontentsline{toc}{section}{References}

\hypertarget{refs}{}
\leavevmode\hypertarget{ref-agrawalSimpleEconomicsCrowdfundinga}{}%
Agrawal, Ajay, Christian Catalini, and Avi Goldfarb. n.d. ``Some Simple
Economics of Crowdfunding,'' 35.

\leavevmode\hypertarget{ref-andreoni_leadership_2006}{}%
Andreoni, James. 2006. ``Leadership Giving in Charitable Fund-Raising.''
\emph{Journal of Public Economic Theory} 8 (1): 1--22.
\url{https://doi.org/10.1111/j.1467-9779.2006.00250.x}.

\leavevmode\hypertarget{ref-augustCompetitionProprietaryOpenSource2021}{}%
August, Terrence, Wei Chen, and Kevin Zhu. 2021. ``Competition Among
Proprietary and Open-Source Software Firms: The Role of Licensing in
Strategic Contribution.'' \emph{Management Science} 67 (5): 3041--66.
\url{https://doi.org/10.1287/mnsc.2020.3674}.

\leavevmode\hypertarget{ref-baker_empirical_1999}{}%
Baker, Michael, A. Abigail Payne, and Michael Smart. 1999. ``An
Empirical Study of Matching Grants: The `Cap on CAP'.'' \emph{Journal of
Public Economics} 72 (2): 269--88.
\url{https://doi.org/10.1016/S0047-2727(98)00092-9}.

\leavevmode\hypertarget{ref-belleflammeCrowdfundingTappingRight2014}{}%
Belleflamme, Paul, Thomas Lambert, and Armin Schwienbacher. 2014.
``Crowdfunding: Tapping the Right Crowd.'' \emph{Journal of Business
Venturing} 29 (5): 585--609.
\url{https://doi.org/10.1016/j.jbusvent.2013.07.003}.

\leavevmode\hypertarget{ref-burtchHiddenCostAccommodating2015}{}%
Burtch, Gordon, Anindya Ghose, and Sunil Wattal. 2015. ``The Hidden Cost
of Accommodating Crowdfunder Privacy Preferences: A Randomized Field
Experiment.'' \emph{Management Science} 61 (5): 949--62.
\url{https://doi.org/10.1287/mnsc.2014.2069}.

\leavevmode\hypertarget{ref-buterin_quadratic_2019}{}%
Buterin, Vitalik. 2019. ``Quadratic Payments: A Primer.''
\url{https://vitalik.ca/general/2019/12/07/quadratic.html}.

\leavevmode\hypertarget{ref-buterinFlexibleDesignFunding2019}{}%
Buterin, Vitalik, Zoë Hitzig, and E. Glen Weyl. 2019. ``A Flexible
Design for Funding Public Goods.'' \emph{Management Science} 65 (11).
INFORMS: 5171--87.

\leavevmode\hypertarget{ref-clarkeMultipartPricingPublic1971}{}%
Clarke, Edward H. 1971. ``Multipart Pricing of Public Goods.''
\emph{Public Choice} 11 (1). Springer: 17--33.

\leavevmode\hypertarget{ref-fehrFairnessRetaliationEconomics2000}{}%
Fehr, Ernst, and Simon Gächter. 2000. ``Fairness and Retaliation: The
Economics of Reciprocity.'' \emph{Journal of Economic Perspectives} 14
(3): 159--81.

\leavevmode\hypertarget{ref-friedmanNoncooperativeEquilibriumSupergames1971}{}%
Friedman, James W. 1971. ``A Non-Cooperative Equilibrium for
Supergames.'' \emph{The Review of Economic Studies} 38 (1). JSTOR:
1--12.

\leavevmode\hypertarget{ref-gobelManagementResearchReciprocity2013}{}%
Göbel, Markus, Rick Vogel, and Christiana Weber. 2013. ``Management
Research on Reciprocity: A Review of the Literature.'' \emph{Business
Research} 6 (1). Springer: 34--53.

\leavevmode\hypertarget{ref-grovesIncentivesTeams1973}{}%
Groves, Theodore. 1973. ``Incentives in Teams.'' \emph{Econometrica:
Journal of the Econometric Society}. JSTOR, 617--31.

\leavevmode\hypertarget{ref-grovesOptimalAllocationPublic1977}{}%
Groves, Theodore, and John Ledyard. 1977. ``Optimal Allocation of Public
Goods: A Solution to the" Free Rider" Problem.'' \emph{Econometrica:
Journal of the Econometric Society}. JSTOR, 783--809.

\leavevmode\hypertarget{ref-huck_matched_2011}{}%
Huck, Steffen, and Imran Rasul. 2011. ``Matched Fundraising: Evidence
from a Natural Field Experiment.'' \emph{Journal of Public Economics},
Charitable giving and fundraising special issue, 95 (5): 351--62.
\url{https://doi.org/10.1016/j.jpubeco.2010.10.005}.

\leavevmode\hypertarget{ref-hyllandEfficientAllocationIndividuals1979}{}%
Hylland, Aanund, and Richard Zeckhauser. 1979. ``The Efficient
Allocation of Individuals to Positions.'' \emph{Journal of Political
Economy} 87 (2). The University of Chicago Press: 293--314.

\leavevmode\hypertarget{ref-lalleyNashEquilibriaQuadratic2019}{}%
Lalley, Steven, and E. Glen Weyl. 2019. ``Nash Equilibria for Quadratic
Voting.'' \emph{Available at SSRN 2488763}.

\leavevmode\hypertarget{ref-meyskens_crowdfunding_2015}{}%
Meyskens, Moriah, and Lacy Bird. 2015. ``Crowdfunding and Value
Creation.'' \emph{Entrepreneurship Research Journal} 5 (2).
\url{https://doi.org/10.1515/erj-2015-0007}.

\leavevmode\hypertarget{ref-mollickDynamicsCrowdfundingExploratory2014}{}%
Mollick, Ethan. 2014. ``The Dynamics of Crowdfunding: An Exploratory
Study.'' \emph{Journal of Business Venturing} 29 (1). Elsevier: 1--16.

\leavevmode\hypertarget{ref-nagleOpenSourceSoftware2019}{}%
Nagle, Frank. 2019. ``Open Source Software and Firm Productivity.''
\emph{Management Science} 65 (3): 1191--1215.
\url{https://doi.org/10.1287/mnsc.2017.2977}.

\leavevmode\hypertarget{ref-nakasaiAnalysisDonationsEclipse2017}{}%
Nakasai, Keitaro, Hideaki Hata, Saya Onoue, and Kenichi Matsumoto. 2017.
``Analysis of Donations in the Eclipse Project.'' In \emph{2017 8th
International Workshop on Empirical Software Engineering in Practice
(IWESEP)}, 18--22. Tokyo, Japan: IEEE.
\url{https://doi.org/10.1109/IWESEP.2017.19}.

\leavevmode\hypertarget{ref-overneyHowNotGet2020}{}%
Overney, Cassandra, Jens Meinicke, Christian Kästner, and Bogdan
Vasilescu. 2020. ``How to Not Get Rich: An Empirical Study of Donations
in Open Source.'' In \emph{Proceedings of the ACM/IEEE 42nd
International Conference on Software Engineering}, 1209--21. Seoul South
Korea: ACM. \url{https://doi.org/10.1145/3377811.3380410}.

\leavevmode\hypertarget{ref-samuelson_pure_1954}{}%
Samuelson, Paul A. 1954. ``The Pure Theory of Public Expenditure.''
\emph{The Review of Economics and Statistics}, 387--89.

\leavevmode\hypertarget{ref-shortResearchCrowdfundingReviewing2017}{}%
Short, Jeremy C., David J. Ketchen, Aaron F. McKenny, Thomas H. Allison,
and R. Duane Ireland. 2017. ``Research on Crowdfunding: Reviewing the
(Very Recent) Past and Celebrating the Present.'' \emph{Entrepreneurship
Theory and Practice} 41 (2): 149--60.
\url{https://doi.org/10.1111/etap.12270}.

\leavevmode\hypertarget{ref-vickreyCounterspeculationAuctionsCompetitive1961}{}%
Vickrey, William. 1961. ``Counterspeculation, Auctions, and Competitive
Sealed Tenders.'' \emph{The Journal of Finance} 16 (1). JSTOR: 8--37.

\leavevmode\hypertarget{ref-vonkroghPromiseResearchOpen2006}{}%
von Krogh, Georg, and Eric von Hippel. 2006. ``The Promise of Research
on Open Source Software.'' \emph{Management Science} 52 (7): 975--83.
\url{https://doi.org/10.1287/mnsc.1060.0560}.

\leavevmode\hypertarget{ref-weylQuadraticVoteBuying2012}{}%
Weyl, E. Glen. 2012. ``Quadratic Vote Buying.'' \emph{Unpublished,
University of Chicago}.

\appendices
\section{Proofs}

\proof{Proof of Proposition \ref{thm:funding} }

As shown in Equation \ref{eq:qfmatch} QF rule requirements can be
decomposed into funds contributed by individual contributors and
matching fund requirements as follows \[
F^{p,\text{QF}}=(\sum \sqrt{c_i})^2=\sum (\sqrt{c_i})^2+2\sum_{i\neq j} \sqrt{c_i}\sqrt{c_j}
\] Dividing each side by \(F^{p,\text{QF}}\) \[
1=\sum \big(\frac{\sqrt{c_i}}{\sum\sqrt{c_i}}\big)^2+\frac{2\sum_{i\neq j} \sqrt{c_i}\sqrt{c_j}}{F^{p,\text{QF}}}
\] Using the definition of \(\alpha_i\) and solving for
\(2\sum_{i\neq j} \sqrt{c_i}\sqrt{c_j}\) \begin{equation}
2\sum_{i\neq j} \sqrt{c_i}\sqrt{c_j}=(1-\sum\alpha_i^2)F^{p,\text{QF}}
\label{eq:requirementsHHI}\end{equation} The LHS of the equation are the
matching fund requirements. Notice that matching fund requirements are
lower the higher is \(\sum\alpha_i^2\). Notice \(\sum\alpha_i^2\)
usually used as the \emph{Herfindahl-Hirschman} HHI concentration index.
This quantity can also be expressed in terms of a variability indicator
as the sampling variance. Using the sampling variance definition: \[
\sigma^2=\frac{1}{n}\sum_i(\alpha_i-\bar{\alpha})^2=\frac{1}{n}\sum_i(\alpha_i^2-2\alpha_i\bar{\alpha}+\bar{\alpha}^2)=\frac{1}{n}\sum_i\alpha_i^2-2n\bar{\alpha}^2+n\bar{\alpha}^2=\frac{1}{n}\sum_i\alpha_i^2-n\bar{\alpha}^2
\] Solving for \(\sum\alpha_i^2\), gives: \[
\sum_i\alpha_i^2=n\sigma^2+n\bar{\alpha}^2
\] Substituting in Equation \ref{eq:requirementsHHI}, gives: \[
2\sum_{i\neq j} \sqrt{c_i}\sqrt{c_j}=(1-n\sigma^2-n\bar{\alpha}^2)F^{p,\text{QF}}
\] \endproof

\proof{Proof of Proposition \ref{thm:correlation} }

The maximum subsidy results from solving the problem: \[
\max_{s_i^p, i\in I, p \in P}2\sum_p\sum_{i\neq j}\sqrt{s_i^pm_is_j^pm_j}+\sum_i\lambda_i(1-\sum_{p}s_i^p)
\] And the first order condition with respect to \(s_i^p\) is \[
2\frac{1}{2}{(s_i^p)}^{-\frac{1}{2}}\sqrt{m_is_j^pm_j}=\lambda_i,  \forall i, p
\] Evaluating this condition for two projects \(p\) and \(p'\) and
dividing each side of the each equation gives: \[
(\frac{s_i^p}{s_i^{p'}})^{-\frac{1}{2}}(\frac{s_j^p}{s_j^{p'}})^{\frac{1}{2}}=1
\] Or alternatively: \[
\frac{s_i^p}{s_i^{p'}}=\frac{s_j^p}{s_j^{p'}}
\] Therefore, the total required subsidy is maximized when the invested
shares across individuals are \emph{perfectly correlated}.

\endproof

\proof{Proof of Proposition \ref{thm:lambdapandconcentration} }

Here we show that, for a given value of \(k\) and \(n\), \(\lambda_p\)is
higher for more equally contributed projects.

Denote \(\alpha_i=\frac{\sqrt{c_i^p}}{\sum_{i}\sqrt{c_i^p}}\) as measure
of the share of project \(p\) contributed by \(i\). We have that \[
\lambda_p \equiv \sum_i\frac{1}{(\frac{1}{k}\frac{1}{\alpha_i^p}+(1-\frac{1}{k}))}\geq\frac{n^2}{\sum_i(\frac{1}{k}\frac{1}{\alpha_i^p}+(1-\frac{1}{k}))}=\frac{n^2}{\frac{1}{k}\sum_i\frac{1}{\alpha_i^p}+n(1-\frac{1}{k})}
\] Where the inequality follows the application of the inequality known
Sedrakyan's inequality, Bergström's inequality, or Titu's lemma (i.e.,
for reals \(a_1,a_2,...,a_n\) and \(b_1, b_2,...,b_n,\) we have
\(\frac{a_1^2}{b_1}+\frac{a_2^2}{b_2}+...+\frac{a_n^2}{b_n}\geq\frac{(a_1+a_2+...+a_n)^2}{b_1+b_2+...+b_n}\)
). The result is a lower bound for \(\lambda_p\). Notice that this value
increases as projects are more equally invested (less concentrated among
contributors). Using Sedrakyan's inequality, and the fact that
\(\sum_i\alpha_i=1\) we also have that
\(\sum_i\frac{1}{\alpha_i}\geq n^2\) , with
\(\sum_i\frac{1}{\alpha_i}= n^2\) for the special case in which
\(\alpha_i=\frac{1}{n}, \forall i\). Indeed,
\(\sum_i\frac{1}{\alpha_i}\) can be viewed as a measure of concentration
(i.e., increases as projects are more unequally contributed). This is
because \(f(x)=\frac{1}{x}\) is a strictly convex function. Therefore
more concentrated invested projects imply lower values of \(\lambda_p\).

\endproof

\proof{Proof of Proposition \ref{thm:collusioninrepetitivegames} }

The proposition follows the application of the result in
\cite{friedmanNoncooperativeEquilibriumSupergames1971} in this specific
case. As an example, consider the static game presented above, and
assume a time discount factor given by the rate \(r\). A collusion can
be sustained using a trigger strategy, of the type ``Invest in the other
party project meanwhile the other party invests back. If the other party
ceases to invest, do not invest in the other party again.'' Note that if
both parties collude following this strategy, this gives individuals a
constant stream of payoffs \(c\) to infinity, with a present value given
by: \(\sum_{i=0}^{\infty} \frac{c}{(1+r)^i}=c+\frac{c}{r}\). If one
party deviates and do not invests, the resulting outcome is given by the
deviation outcome in the present, plus 0 in the future.(i.e.,:
\(c(\frac{1+2\sqrt{2}}{2})\)). This implies that the collusion will be
maintained as long as \(c+\frac{c}{r}\geq c(\frac{1+2\sqrt{2}}{2})\),
which implies, in this case, that the collusion will be maintained while
\(r\leq \frac{2}{2\sqrt{2}-1}\approx 1.09=109\). Since it is reasonable
to expect \(r\) to remain below 109, it follows that in this particular
game there are incentives to maintain the collusion indefinitely.\\
\endproof

\proof{Proof of Proposition \ref{thmcollusionn}}

Assuming that only a fraction \(\alpha\) of the \(n\) individuals
contribute \(\frac{c}{n}\), the total amount contributed by backers is
\(\alpha c\) , and the target QF matching to resulting from the
mechanism would be \(\big(\alpha n \sqrt{\frac{c}{n}}\big)^2\) in the
absence of budget limits to the fund. If there are contrains on the pool
of matching funds, the actual amounts to be received by the project, as
defined in Equation \ref{eq:actualfunds} are \[
F^p=\frac{1}{k}\bigg[\big(\alpha n \sqrt{\frac{c}{n}}\big)^2-\alpha c\bigg]+\alpha c=c\bigg[\ \frac{1}{k}\big(\alpha^2n-\alpha\big)+\alpha\bigg]
\] The return of investing \(c\) in this strategy will be positive if
\(F^p-c>0\), or \[
\bigg[\ \frac{1}{k}\big(\alpha^2n-\alpha\big)+\alpha\bigg]-1>0
\] Which is equalivalent to \[
\frac{n}{k}\alpha^2+(1-\frac{1}{k})\alpha-1>0
\] Solving the quadratic equation yields that \[
\alpha^{**}>\frac{k\bigg((1-\frac{1}{k})+\sqrt{(1-\frac{1}{k})^2+4(\frac{n}{k})}\bigg)}{2n}
\] Note that if \(k \xrightarrow{} 1\) , then
\(\alpha^{**} \xrightarrow{} {\frac{1}{\sqrt{n}}}\) , which is Equation
\ref{eq:alpha}. Also if , \(k \xrightarrow{} +\infty\) then
\(\alpha^{**} \xrightarrow{} {+\infty}\).

\endproof
\section{Tables and figures}

\begin{figure}
\hypertarget{fig:round8_k_part1}{%
\centering
\includegraphics[width=4.47917in,height=4.375in]{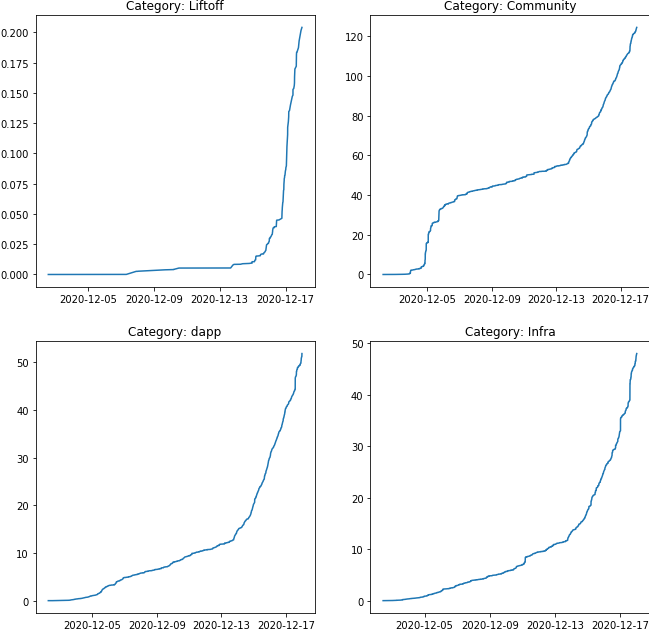}
\caption{Evolution of budget scale constant \(k\) during
round8}\label{fig:round8_k_part1}
}
\end{figure}

\begin{figure}
\hypertarget{fig:round8_k_part2}{%
\centering
\includegraphics[width=4.47917in,height=2.35417in]{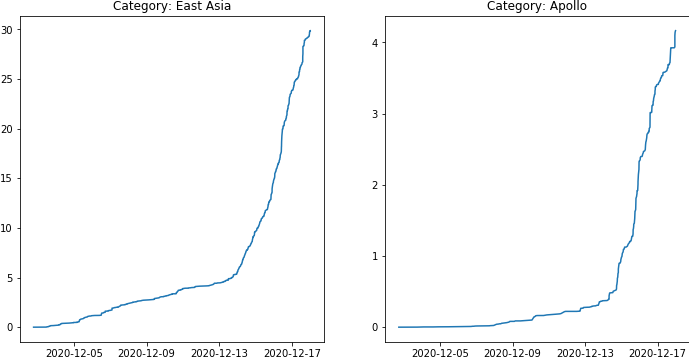}
\caption{Evolution of budget scale constant \(k\) during round
8}\label{fig:round8_k_part2}
}
\end{figure}

\begin{center}

\begin{table}[!htbp]

\caption{Gitcoin Round 8 Descriptive Statistics\label{tbl:sumstats8}}

\small

\begin{tabular}{lccccc}
\toprule
Category & Variable & N & Mean & Std. Dev. & Median \\ \hline
All & $c_i$ & 18392 & 29.632 & 482.553 & 4.470 \\
\textit{(projects: 444 contributors:   4953)} & $\sqrt{c_i}$ & 18392 & 2.625 & 4.769 & 2.114 \\
Dapp & $c_i$ & 6317 & 19.466 & 379.871 & 3.800 \\
\textit{(projects: 182 contributors:   2608)} & $\sqrt{c_i}$ & 6317 & 2.385 & 3.712 & 1.949 \\
Community & $c_i$ & 7311 & 37.862 & 632.955 & 4.250 \\
\textit{(projects: 177 contributors:   2854)} & $\sqrt{c_i}$ & 7311 & 2.697 & 5.531 & 2.062 \\
Infra & $c_i$ & 4295 & 25.742 & 301.379 & 4.410 \\
\textit{(projects: 55 contributors:   1431)} & $\sqrt{c_i}$ & 4295 & 2.551 & 4.386 & 2.100 \\
East-Asia & $c_i$ & 1293 & 16.619 & 88.176 & 3.800 \\
\textit{(projects: 38 contributors:   811)} & $\sqrt{c_i}$ & 1293 & 2.662 & 3.088 & 1.949 \\
Lift-off & $c_i$ & 137 & 28.540 & 245.225 & 0.998 \\
\textit{(projects: 10 contributors:   119)} & $\sqrt{c_i}$ & 137 & 1.874 & 5.021 & 0.999 \\
Apollo & $c_i$ & 332 & 27.908 & 198.974 & 2.000 \\
\textit{(projects: 20 contributors:   287)} & $\sqrt{c_i}$ & 332 & 2.608 & 4.601 & 1.414 \\ \hline
\end{tabular}

\vspace{1ex}

{\raggedright Note: This table reports summary statistics on the total backer contributions per project \par}

\end{table}

\end{center}

\begin{figure}
\hypertarget{fig:histcip_round8a}{%
\centering
\includegraphics[width=5.20833in,height=5.58333in]{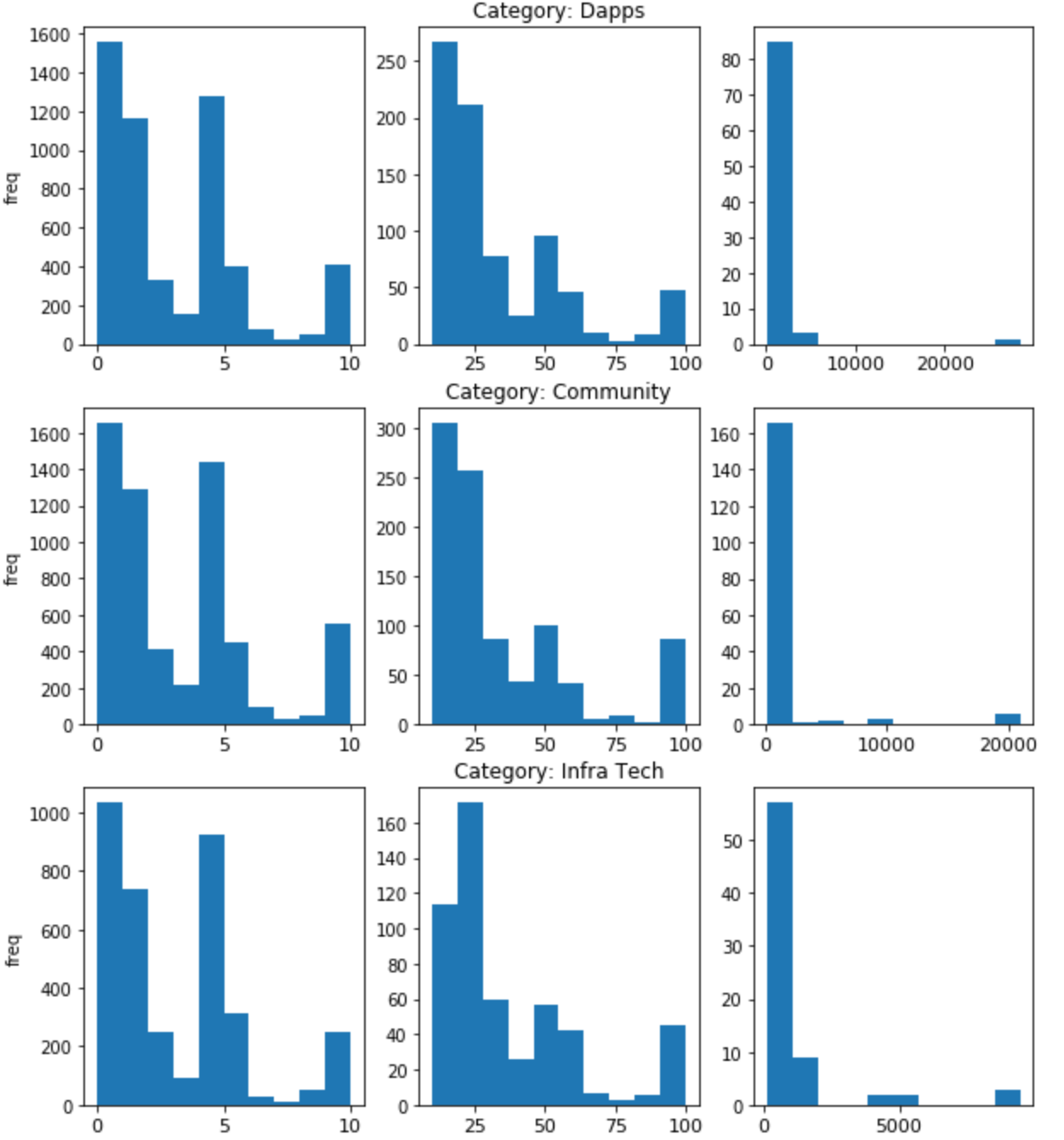}
\caption{Split range histograms of total individual contributions to
projects \(c_i^p\) by category. Round 8}\label{fig:histcip_round8a}
}
\end{figure}

\begin{figure}
\hypertarget{fig:histcip_round8b}{%
\centering
\includegraphics[width=5.20833in,height=5.58333in]{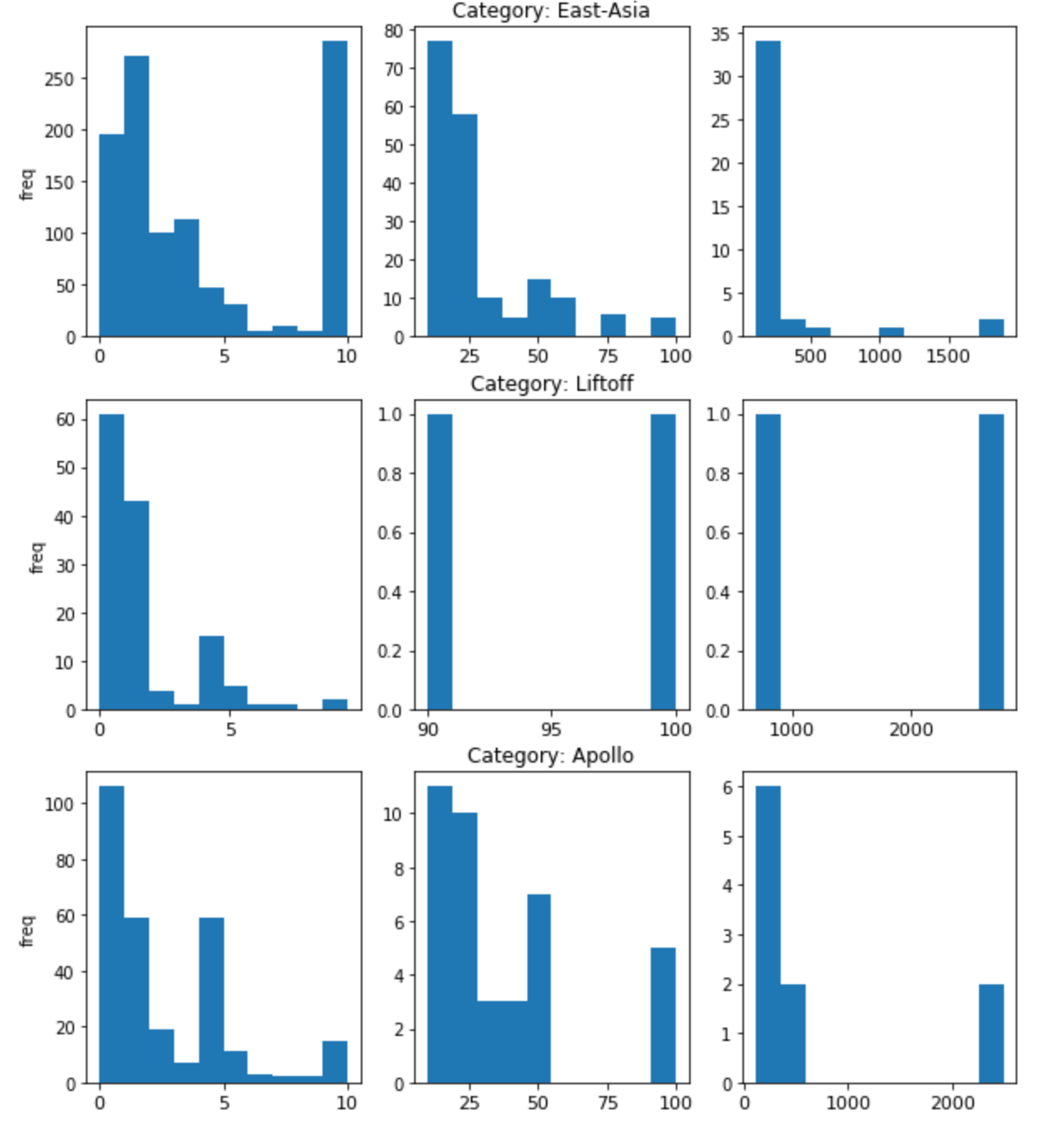}
\caption{Split range histograms of total individual contributions to
projects \(c_i^p\) by category. Round 8}\label{fig:histcip_round8b}
}
\end{figure}

\endappendices

\bibliographystyle{chicago}

\bibliography{Bibliography-MM-MC}

\end{document}